\documentclass[apj,numberedappendix,revtex4]{emulateapj}
\usepackage{amsmath}
\usepackage{natbib}
\usepackage{graphicx}
\usepackage{color}

\DeclareGraphicsExtensions{.jpg,.pdf,.png,.eps,.ps}

\newcommand{\msun}{\ensuremath{\mathrm{M}_\odot}}
\newcommand{\chandra}{\emph{Chandra}}
\newcommand{\covm}{\ensuremath{\Psi}}
\newcommand{\Yx}{\mbox{$Y_\mathrm{X}$}}
\newcommand{\subYx}{\ensuremath{\mathrm{Yx}}}
\newcommand{\omm}{\ensuremath{\Omega_m}}
\newcommand{\seight}{\ensuremath{\sigma_8 \left(\omm/0.27\right)^{0.3}}}
\newcommand{\falsenorm}{\ensuremath{\alpha_\mathrm{field}}}
\newcommand{\falseslope}{\ensuremath{\beta_\mathrm{field}}}
\newcommand{\fsf}{\ensuremath{\gamma_\mathrm{field}}}
\newcommand{\summnu}{\ensuremath{\Sigma m_\nu}}
\newcommand{\neff}{\ensuremath{N_\mathrm{eff}}}
\newcommand{\spitzer}{{\sl Spitzer}}
\newcommand{\planck}{{\sl Planck}}
\newcommand{\wmap}{{\sl WMAP}}
\newcommand{\um}{$\mu$m}
\newcommand{\degs}{\ensuremath{\mathrm{deg}^2}}
\newcommand{\yxm}{\mbox{\Yx\ - $M$}}
\newcommand{\zetam}{\mbox{$\zeta$ - $M$}}
\newcommand{\lcdm}{\mbox{$\Lambda$CDM}}
\newcommand{\wcdm}{\mbox{$w$CDM}}
\newcommand{\sptcl}{SPT\mbox{$_{\mbox{\scriptsize CL}}$}}
\newcommand{\actcl}{ACT\mbox{$_{\mbox{\scriptsize CL}}$}}
\newcommand{\myvector}[1]{\textbf{#1}}
\newcommand{\overdensityradius}{\ensuremath{r}}

\newcommand{\nclust}{377} 
\newcommand{\nclustx}{82} 
\newcommand{\constrainthsptclsigmaeight}{\ensuremath{0.784 \pm 0.039}} 
\newcommand{\constrainthsptclomegam}{\ensuremath{0.289 \pm 0.042}} 
\newcommand{\constrainthsptclseight}{\ensuremath{0.797 \pm 0.031}} 
\newcommand{\mnuplanckbaoresult}{\ensuremath{0.14 \pm 0.08}} 
\newcommand{\nnuresultsptclplanck}{\ensuremath{3.25 \pm 0.23}} 
\newcommand{\mnuresultsptclplancknnufree}{\ensuremath{0.39 \pm 0.20}} 
\newcommand{\nnuresultsptclplanckfracimprovement}{\ensuremath{1.3}} 
\newcommand{\mnuresultsptclplancknnufreefracimprovement}{\ensuremath{1.6}} 
\newcommand{\nnuplanckbaoresult}{\ensuremath{3.28 \pm 0.20}} 
\newcommand{\mnuplanckbaoresultnnufree}{\ensuremath{0.18 \pm 0.09}} 
\newcommand{\nnuallresult}{\ensuremath{3.43 \pm 0.16}} 
\newcommand{\mnuallresultnnufree}{\ensuremath{0.16 \pm 0.08}} 
\newcommand{\nnuallsignificance}{\ensuremath{2.3\sigma}} 
\newcommand{\wresultsptclw}{\ensuremath{-1.28 \pm 0.31}} 
\newcommand{\wresultsptclomegal}{\ensuremath{0.738 \pm 0.046}} 
\newcommand{\wresultplancksptclw}{\ensuremath{-1.04 \pm 0.17}} 
\newcommand{\wresultplancksptclsigmaeight}{\ensuremath{0.803 \pm 0.045}} 
\newcommand{\wresulthbaosnesptclw}{\ensuremath{-1.08 \pm 0.07}} 
\newcommand{\wresulthbaosnefactor}{\ensuremath{37\%}} 
\newcommand{\wresultallw}{\ensuremath{-1.023 \pm 0.042}} 
\newcommand{\wresultallwnosptcl}{\ensuremath{-1.062 \pm 0.048}} 
\newcommand{\wresultallfactor}{\ensuremath{14\%}} 

\def\Berkeley{1}
\def\McGill{2}
\def\AAUChicago{3}
\def\KICPChicago{4}
\def\FNAL{5}
\def\PhysicsUChicago{6}
\def\Argonne{7}
\def\StanfordKPAC{8}
\def\StanfordPhys{9}
\def\SLAC{10}
\def\AIfA{11}
\def\CfA{12}
\def\MIT{13}
\def\Harvard{14}
\def\Munich{15}
\def\ExcellenceCluster{16}
\def\Miss{17}
\def\EFIChicago{18}
\def\NIST{19}
\def\PUC{20}
\def\Caltech{21}
\def\CIFAR{22}
\def\Melbourne{23}
\def\illast{24}
\def\illphy{25}
\def\Huntingdon{26}
\def\MPE{27}
\def\UFlorida{28}
\def\Colorado{29}
\def\UMon{30}
\def\KIPAC{31}
\def\LeidenObservatory{32}
\def\UChicago{33}
\def\Davis{34}
\def\LBNL{35}
\def\StonyBrook{36}
\def\DARK{37}
\def\Arizona{38}
\def\Michigan{39}
\def\Minnesota{40}
\def\STScI{41}
\def\CaseWestern{42}
\def\ArtInstChicago{43}
\def\KASI{44}
\def\LLNL{45}
\def\Dunlap{46}
\def\Toronto{47}
\def\CTIO{48}

\begin{document}

 \title{Cosmological Constraints from Galaxy Clusters \\ in the 2500 square-degree SPT-SZ Survey}

 \slugcomment{Submitted to the Astrophysical Journal}
 

 \author{
  T.~de~Haan\altaffilmark{\Berkeley, \McGill},
  B.~A.~Benson\altaffilmark{\AAUChicago,\KICPChicago,\FNAL},
  L.~E.~Bleem\altaffilmark{\KICPChicago,\PhysicsUChicago,\Argonne},
  S.~W.~Allen\altaffilmark{\StanfordKPAC,\StanfordPhys,\SLAC},
  D.~E.~Applegate\altaffilmark{\AIfA},
  M.~L.~N.~Ashby\altaffilmark{\CfA},
  M.~Bautz\altaffilmark{\MIT},
  M.~Bayliss\altaffilmark{\CfA,\Harvard},
  S.~Bocquet\altaffilmark{\Munich,\ExcellenceCluster,\KICPChicago,\Argonne},
  M.~Brodwin\altaffilmark{\Miss},
  J.~E.~Carlstrom\altaffilmark{\AAUChicago,\KICPChicago,\PhysicsUChicago,\Argonne,\EFIChicago},
  C.~L.~Chang\altaffilmark{\AAUChicago,\KICPChicago,\Argonne},
  I.~Chiu\altaffilmark{\Munich,\ExcellenceCluster},
  H-M.~Cho\altaffilmark{\NIST},
  A.~Clocchiatti\altaffilmark{\PUC},
  T.~M.~Crawford\altaffilmark{\AAUChicago,\KICPChicago}
  A.~T.~Crites\altaffilmark{\AAUChicago,\KICPChicago,\Caltech},
  S.~Desai\altaffilmark{\Munich,\ExcellenceCluster},
  J.~P.~Dietrich\altaffilmark{\Munich,\ExcellenceCluster},
  M.~A.~Dobbs\altaffilmark{\McGill,\CIFAR},
  A.~N.~Doucouliagos\altaffilmark{\Melbourne},
  R.~J.~Foley\altaffilmark{\illast,\illphy},
  W.~R.~Forman\altaffilmark{\CfA},
  G.~P.~Garmire\altaffilmark{\Huntingdon},
  E.~M.~George\altaffilmark{\Berkeley,\MPE},
  M.~D.~Gladders\altaffilmark{\AAUChicago,\KICPChicago},
  A.~H.~Gonzalez\altaffilmark{\UFlorida},
  N.~Gupta\altaffilmark{\Munich,\ExcellenceCluster},
  N.~W.~Halverson\altaffilmark{\Colorado},
  J.~Hlavacek-Larrondo\altaffilmark{\UMon,\KIPAC,\StanfordPhys},
  H.~Hoekstra\altaffilmark{\LeidenObservatory},
  G.~P.~Holder\altaffilmark{\McGill},
  W.~L.~Holzapfel\altaffilmark{\Berkeley},
  Z.~Hou\altaffilmark{\KICPChicago,\PhysicsUChicago},
  J.~D.~Hrubes\altaffilmark{\UChicago},
  N.~Huang\altaffilmark{\Berkeley},
  C.~Jones\altaffilmark{\CfA},
  R.~Keisler\altaffilmark{\KICPChicago,\StanfordKPAC,\StanfordPhys,\PhysicsUChicago},
  L.~Knox\altaffilmark{\Davis},
  A.~T.~Lee\altaffilmark{\Berkeley,\LBNL},
  E.~M.~Leitch\altaffilmark{\AAUChicago,\KICPChicago},
  A.~von~der~Linden\altaffilmark{\StonyBrook,\DARK,\StanfordKPAC,\StanfordPhys},
  D.~Luong-Van\altaffilmark{\UChicago},
  A.~Mantz\altaffilmark{\KICPChicago,\StanfordKPAC,\StanfordPhys},
  D.~P.~Marrone\altaffilmark{\Arizona},
  M.~McDonald\altaffilmark{\MIT},
  J.~J.~McMahon\altaffilmark{\Michigan},
  S.~S.~Meyer\altaffilmark{\AAUChicago,\KICPChicago,\PhysicsUChicago,\EFIChicago},
  L.~M.~Mocanu\altaffilmark{\AAUChicago,\KICPChicago},
  J.~J.~Mohr\altaffilmark{\MPE,\Munich,\ExcellenceCluster},
  S.~S.~Murray\altaffilmark{\CfA},
  S.~Padin\altaffilmark{\AAUChicago,\KICPChicago,\Caltech},
  C.~Pryke\altaffilmark{\Minnesota},
  D.~Rapetti\altaffilmark{\Munich, \ExcellenceCluster},
  C.~L.~Reichardt\altaffilmark{\Melbourne},
  A.~Rest\altaffilmark{\STScI},
  J.~Ruel\altaffilmark{\Harvard}, 
  J.~E.~Ruhl\altaffilmark{\CaseWestern},
  B.~R.~Saliwanchik\altaffilmark{\CaseWestern},
  A.~Saro\altaffilmark{\Munich, \ExcellenceCluster},
  J.~T.~Sayre\altaffilmark{\Colorado},
  K.~K.~Schaffer\altaffilmark{\KICPChicago,\EFIChicago,\ArtInstChicago},
  T.~Schrabback\altaffilmark{\AIfA},
  E.~Shirokoff\altaffilmark{\AAUChicago,\KICPChicago},
  J.~Song\altaffilmark{\Michigan,\KASI},
  H.~G.~Spieler\altaffilmark{\LBNL},
  B.~Stalder\altaffilmark{\CfA},
  S.~A.~Stanford\altaffilmark{\Davis,\LLNL},
  Z.~Staniszewski\altaffilmark{\CaseWestern},
  A.~A.~Stark\altaffilmark{\CfA},
  K.~T.~Story\altaffilmark{\StanfordKPAC,\StanfordPhys},
  C.~W.~Stubbs\altaffilmark{\CfA,\Harvard},
  K.~Vanderlinde\altaffilmark{\Dunlap,\Toronto},
  J.~D.~Vieira\altaffilmark{\illast,\illphy},
  A. Vikhlinin\altaffilmark{\CfA},
  R.~Williamson\altaffilmark{\AAUChicago,\KICPChicago,\Caltech},
  and
  A.~Zenteno\altaffilmark{\CTIO}
}

\altaffiltext{\Berkeley}{Department of Physics, University of California, Berkeley, CA, USA 94720}
\altaffiltext{\McGill}{Department of Physics, McGill University, Montreal, Quebec H3A 2T8, Canada}
\altaffiltext{\AAUChicago}{Department of Astronomy and Astrophysics, University of Chicago, Chicago, IL, USA 60637}
\altaffiltext{\KICPChicago}{Kavli Institute for Cosmological Physics, University of Chicago, Chicago, IL, USA 60637}
\altaffiltext{\FNAL}{Fermi National Accelerator Laboratory, Batavia, IL 60510-0500, USA}
\altaffiltext{\PhysicsUChicago}{Department of Physics, University of Chicago, Chicago, IL, USA 60637} 
\altaffiltext{\Argonne}{Argonne National Laboratory, Argonne, IL, USA 60439}
\altaffiltext{\StanfordKPAC}{Kavli Institute for Particle Astrophysics and Cosmology, Stanford University, 452 Lomita Mall, Stanford, CA 94305} 
\altaffiltext{\StanfordPhys}{Department of Physics, Stanford University, 382 Via Pueblo Mall, Stanford, CA 94305}
\altaffiltext{\SLAC}{SLAC National Accelerator Laboratory, 2575 Sand Hill Road, Menlo Park, CA 94025}
\altaffiltext{\AIfA}{Argelander-Institut f\"ur Astronomie, Auf dem H\"ugel 71, D-53121 Bonn, Germany}
\altaffiltext{\CfA}{Harvard-Smithsonian Center for Astrophysics, Cambridge, MA, USA 02138}
\altaffiltext{\MIT}{Kavli Institute for Astrophysics and Space Research, Massachusetts Institute of Technology, 77 Massachusetts Avenue, Cambridge, MA 02139}
\altaffiltext{\Harvard}{Department of Physics, Harvard University, 17 Oxford Street, Cambridge, MA 02138}
\altaffiltext{\Munich}{Department of Physics, Ludwig-Maximilians-Universit\"{a}t, Scheinerstr.\ 1, 81679 Munich, Germany}
\altaffiltext{\ExcellenceCluster}{Excellence Cluster Universe, Boltzmannstr.\ 2, 85748 Garching, Germany}
\altaffiltext{\Miss}{Department of Physics and Astronomy, University of Missouri, 5110 Rockhill Road, Kansas City, MO 64110}
\altaffiltext{\EFIChicago}{Enrico Fermi Institute, University of Chicago, Chicago, IL, USA 60637}
\altaffiltext{\NIST}{NIST Quantum Devices Group, Boulder, CO, USA 80305}
\altaffiltext{\PUC}{Departamento de Astronomia y Astrosifica, Pontificia Universidad Catolica,Chile}
\altaffiltext{\Caltech}{California Institute of Technology, Pasadena, CA, USA 91125}
\altaffiltext{\CIFAR}{Canadian Institute for Advanced Research, CIFAR Program in Cosmology and Gravity, Toronto, ON, M5G 1Z8, Canada}
\altaffiltext{\Melbourne}{School of Physics, University of Melbourne, Parkville, VIC 3010, Australia}
\altaffiltext{\illast}{Astronomy Department, University of Illinois at Urbana-Champaign, 1002 W.\ Green Street, Urbana, IL 61801, USA}
\altaffiltext{\illphy}{Department of Physics, University of Illinois Urbana-Champaign, 1110 W.\ Green Street, Urbana, IL 61801, USA}
\altaffiltext{\Huntingdon}{Huntingdon Institute for X-ray Astronomy, LLC}
\altaffiltext{\MPE}{Max Planck Institute for Extraterrestrial Physics, Giessenbachstr.\ 1, 85748 Garching, Germany}
\altaffiltext{\UFlorida}{Department of Astronomy, University of Florida, Gainesville, FL 32611}
\altaffiltext{\Colorado}{Department of Astrophysical and Planetary Sciences and Department of Physics, University of Colorado, Boulder, CO, USA 80309}
\altaffiltext{\UMon}{Department of Physics, Universit\'e de Montr\'eal, Montrea, Quebec H3T 1J4, Canada}
\altaffiltext{\KIPAC}{Kavli Institute for Particle Astrophysics and Cosmology, Stanford University, 452 Lomita Mall, Stanford, CA 94305-4085, USA}
\altaffiltext{\LeidenObservatory}{Leiden Observatory, Leiden University, Niels Bohrweg 2, 2333 CA, Leiden, the Netherlands}
\altaffiltext{\UChicago}{University of Chicago, Chicago, IL, USA 60637} 
\altaffiltext{\Davis}{Department of Physics, University of California, Davis, CA, USA 95616}
\altaffiltext{\LBNL}{Physics Division, Lawrence Berkeley National Laboratory, Berkeley, CA, USA 94720}
\altaffiltext{\StonyBrook}{Department of Physics and Astronomy, Stony Brook University, Stony Brook, NY 11794, USA}
\altaffiltext{\DARK}{Dark Cosmology Centre, Niels Bohr Institute, University of Copenhagen Juliane Maries Vej 30, 2100 Copenhagen, Denmark}
\altaffiltext{\Arizona}{Steward Observatory, University of Arizona, 933 North Cherry Avenue, Tucson, AZ 85721}
\altaffiltext{\Michigan}{Department of Physics, University of Michigan, Ann  Arbor, MI, USA 48109}
\altaffiltext{\Minnesota}{Department of Physics, University of Minnesota, Minneapolis, MN, USA 55455}
\altaffiltext{\STScI}{Space Telescope Science Institute, 3700 San Martin Dr., Baltimore, MD 21218}
\altaffiltext{\CaseWestern}{Physics Department, Center for Education and Research in Cosmology and Astrophysics, Case Western Reserve University,Cleveland, OH, USA 44106}
\altaffiltext{\ArtInstChicago}{Liberal Arts Department, School of the Art Institute of Chicago, Chicago, IL, USA 60603}
\altaffiltext{\KASI}{Korea Astronomy and Space Science Institute, Daejeon 305-348, Republic of Korea}
\altaffiltext{\LLNL}{Institute of Geophysics and Planetary Physics, Lawrence Livermore National Laboratory, Livermore, CA 94551}
\altaffiltext{\Dunlap}{Dunlap Institute for Astronomy \& Astrophysics, University of Toronto, 50 St George St, Toronto, ON, M5S 3H4, Canada}
\altaffiltext{\Toronto}{Department of Astronomy \& Astrophysics, University of Toronto, 50 St George St, Toronto, ON, M5S 3H4, Canada}
\altaffiltext{\CTIO}{Cerro Tololo Inter-American Observatory, Casilla 603, La Serena, Chile}

\begin{abstract}
  We present cosmological parameter constraints obtained from galaxy clusters identified by their Sunyaev-Zel'dovich effect signature in the 2500 square degree South Pole Telescope Sunyaev Zel'dovich (SPT-SZ) survey. 
  We consider the \nclust\ cluster candidates identified at $z>0.25$ with a 
  detection significance greater than five, corresponding to the $95\%$ purity threshold for the survey. 
  We compute constraints on cosmological models using the measured cluster abundance as a function of mass and redshift. 
  We include additional constraints from multi-wavelength observations, including \chandra\ X-ray data for \nclustx\ clusters
  and a weak lensing-based prior on the normalization of the mass-observable scaling relations. 
  We introduce a numerical technique to efficiently compute the cluster likelihood in the presence of an arbitrary number 
  of mass proxies, each of which relate to the underlying cluster mass with a power-law scaling relation and contain scatter, of which a log-normal component may be correlated between mass proxies.
  Assuming a spatially flat
  \lcdm\ cosmology, where the species-summed neutrino mass has the minimum allowed value ($\summnu = 0.06~\mathrm{eV}$) from neutrino oscillation experiments, 
  we combine the cluster data with a prior on $H_0$ and
  find $\sigma_8 = \constrainthsptclsigmaeight$ and $\omm = \constrainthsptclomegam$,
  with the parameter combination $\seight = \constrainthsptclseight$. 
  These results are in good agreement with constraints from the cosmic microwave background (CMB) from SPT, \wmap, and \planck, as well as with constraints from other cluster datasets.  
  Adding the sum of the neutrino masses as a free parameter, we find $\summnu = \mnuplanckbaoresult~\mathrm{eV}$ 
  when combining the SPT cluster data with \planck\ CMB data and baryon acoustic oscillation (BAO) data, consistent with the minimum allowed value.
  With this dataset, when we simultaneously free \summnu\ and the effective number of relativistic species (\neff), we find $\neff = \nnuplanckbaoresult$ and $\summnu = \mnuplanckbaoresultnnufree~\mathrm{eV}$.
  Finally, we consider a cosmology where \summnu\ and \neff\ are fixed to the \lcdm\ values, but the dark energy equation of state parameter $w$ is free.  Using the SPT cluster data in combination with an $H_0$ prior, 
  we measure $w = \wresultsptclw$, a constraint consistent with the \lcdm\ cosmological model and derived from 
  the combination of growth of structure and geometry. 
  When combined with primarily geometrical constraints from \planck\ CMB, $H_0$, BAO and SNe, adding the SPT cluster data improves the $w$ constraint from the geometrical data alone by \wresultallfactor, to $w = \wresultallw$.
\end{abstract}
\keywords{cosmology: observations --  galaxies: clusters: general}
 
\section{Introduction}
\label{sec:intro}

\setcounter{footnote}{0}
 
Galaxy clusters trace extreme peaks in the matter density field on megaparsec scales.
The abundance of these peaks as a function of mass and redshift is highly sensitive
to the matter density and the growth of structure. As this abundance can be predicted
with sufficient accuracy for a given cosmology \citep{tinker08, bhattacharya10, holder01a}, even modest
measurements of cluster abundance can yield powerful cosmological constraints. 
These constraints are particularly powerful when combined with or compared to independent constraints
from probes such as the power spectrum of the cosmic microwave background (CMB)
and baryon acoustic oscillations (BAO). Taken together, growth-based and geometrical probes can place 
significantly tighter constraints on parameters such as the equation of state of dark energy than either
one independently, because of 
nearly orthogonal parameter degeneracies. Considered independently, these constraints provide a 
consistency test of 
the dark energy paradigm and the validity of general relativity \citep{ishak06, zhan09, mortonson09, mortonson10b, rapetti09, rapetti10, acquaviva10, vanderveld12}.

In recent years, constraints on cosmological parameters from cluster abundance measurements have advanced
significantly using cluster samples selected at
X-ray \citep[e.g.,][]{vikhlinin09, mantz10a, mantz15}, 
optical \citep[e.g.,][]{rozo10}, 
and millimeter \citep[e.g.,][]{reichardt13, planck13-20, hasselfield13, bocquet15, planck15-24} wavelengths.
The predominant millimeter-wave (mm-wave) 
signal from clusters arises from the thermal Sunyaev-Zel'dovich (tSZ) effect \citep{sunyaev72},
i.e., the scattering of CMB photons by hot electrons in 
the intra-cluster medium (ICM). The surface brightness of the tSZ effect is  
redshift-independent, allowing high-resolution mm-wave surveys to obtain nearly 
mass-limited samples of clusters to arbitrarily high redshift. The ability to cleanly select 
clusters out to the redshift at which dark energy begins to contribute significantly to the energy budget
of the universe ($z \sim 1$) is particularly important for constraints on the dark energy equation of
state and tests of the dark energy paradigm. Cluster surveys using high-resolution tSZ data 
are uniquely positioned to deliver such constraints \citep{carlstrom02}.

The primary limitation to cosmological constraints from current cluster surveys at all wavelengths
is an imperfect understanding of the relationship between the quantity that can be predicted from theory or 
simulations (the cluster mass or the height of the associated density peak) and the observable property of
the cluster that is used as a proxy for this quantity. 
The mass proxy can be the observable quantity used to construct the sample; it can also include 
observables from follow-up observations, often at different wavelengths than the cluster selection observable.
Since the cluster abundance is an extremely steep function of mass, misestimation of the relation between the mass proxy and the true cluster mass can lead to significant biases on the resulting cosmological parameter constraints. 

Different cluster mass proxies have distinct advantages and disadvantages related to the 
accuracy and precision with which they trace the true cluster mass and the expense of obtaining 
the data required to construct them. In terms of ultimate accuracy, or absence of bias, the current gold
standard mass proxy is derived from measurements of weak gravitational lensing of background galaxies
by clusters (see \citealt{vonderlinden14b} for a discussion).  
Optical mass measurements from weak gravitational lensing
are observationally expensive to obtain at high redshift, and the
scatter on individual cluster mass estimates is large.
Gravitational lensing of the CMB by clusters 
is a promising future avenue for mass estimation---e.g., \citealt{madhavacheril14,baxter14,planck15-24}.
Cluster velocity dispersions, generally obtained through spectroscopic observations of tens
of cluster member galaxies, are also mostly unaffected by complex ICM physics 
but are expensive observationally and 
have large scatter \citep[e.g.,][]{evrard08,saro13a,sifon13,ruel14}, 
as well as uncertain velocity bias \citep[e.g.,][]{munari13}.
Where weak lensing measurements provide high accuracy, measurements of the gas mass $M_\mathrm{gas}$, and/or of 
the integrated cluster pressure $Y$ provide
high precision.
Estimated from tSZ or X-ray data, integrated cluster pressure 
is predicted and measured to track cluster mass with low scatter 
\citep[e.g.,][]{motl05,nagai07,stanek10,kravtsov06a},
but its relation to true cluster mass can be complicated by non-thermal
pressure support in clusters \citep[e.g.,][]{evrard96,nagai07,yu15}. Furthermore, robust $Y$ estimates require either 
relatively high-quality X-ray data (to provide deprojected temperature and density profiles), or tSZ measurements with 
accurate information on all cluster scales (sub-arcminute
scales to the virial radius). 
Mass proxies built from the same data used to select clusters,  
such as optical richness, X-ray luminosity, and tSZ detection significance, come at no extra cost
but can demonstrate high scatter and
require external calibration data to tie them robustly to the true cluster mass.

 In this work, we use a sample of clusters derived from the SPT-SZ survey, a three-band mm-wave survey 
of 2500 \degs\ of the southern sky conducted with the South Pole Telescope \citep[SPT,][]{carlstrom11}.
The cluster selection method, redshift determination, and sample characteristics are described in detail in  
\citet[][hereafter B15]{bleem15b}. We use the mm-wave and redshift information from B15
in conjunction with targeted X-ray follow-up observations from \emph{Chandra} to obtain cosmological constraints. 
The mass proxy we use is the tSZ detection significance $\xi$, calibrated using X-ray integrated pressure \Yx, 
which is in turn tied to true cluster mass using optical weak lensing (WL) measurements.

The paper is organized as follows. In \S\ref{sec:szdata}, we briefly describe the cluster sample, 
including the mm-wave data and analysis methods that went into producing the sample. 
We summarize the optical/infrared data and redshift estimation in \S\ref{sec:redshifts}, 
and the X-ray data and analysis methods in \S\ref{sec:xray}.
The cosmological analysis methods are described in \S\ref{sec:cosmo}, and we 
present the cosmological parameter constraints in \S\ref{sec:constraints}. We compare the 
cosmological constraints from this work to those from other cluster surveys in \S\ref{sec:compare},   
and we conclude in \S\ref{sec:concl}.   

When parameter constraints are reported, the best-fit value and uncertainties correspond to the mean
and standard deviation of the posterior distribution.
Cluster masses, denoted with $M_{\Delta}$, refer to spherical overdensities for which the enclosed density is equal to $\Delta$ times the critical density, $\rho_c$. Similarly, $\overdensityradius_{\Delta}$ refers to the associated radius such that $M_{\Delta} = \frac{4}{3} \pi \overdensityradius_\Delta^3 \Delta \rho_c(z)$.

\section{SZ Data and Methods}
\label{sec:szdata}

The cluster sample used in this work is a subset of that previously presented in B15. We choose the clusters with
redshift $z>0.25$ and detection significance $\xi > 5$. The significance cut was chosen such that 
the resulting catalog has high ($\sim 95\%$) purity, and the redshift cut allows for a nearly redshift-independent selection function \citep[][hereafter V10]{vanderlinde10}.
Our strategy for tying the cluster abundance measurement to the cosmologically predicted halo mass function is to calibrate the SZ-mass scaling relation using \chandra\, X-ray measurements. The X-ray scaling relation is taken from \citet[][hereafter V09]{vikhlinin09b, vikhlinin09}, though we modify the
overall normalization of this relation in \S\ref{sec:wlprior} to be consistent with more recent weak lensing measurements from \citet[][hereafter H15]{hoekstra15} and the Weighing the Giants (WtG) project \citep{applegate14, vonderlinden14a, mantz15}.

\subsection{Sample of SZ Cluster Candidates}
\label{sec:sz_sample}

The SPT is a 10-m telescope located at the geographic South Pole. With a 1~degree field of view and $\sim1$~arcminute resolution, it was designed to rapidly map large areas of sky while being well matched to the angular size of high-redshift clusters. The SPT-SZ camera operated from 2007 through 2011 and consisted of a 960-element, photon-noise-limited bolometer 
array observing in three frequency bands centered at 95, 150, and 220~GHz, though this work uses only the first two bands.  
The observation strategy and analysis for the 2500-square-degree SPT-SZ survey
are described in many previous SPT papers (e.g. \citealt{schaffer11}), and the
analysis specific to obtaining the cluster sample is described in B15. 
Briefly, the majority of the 2500 square-degree survey was performed using the following scan strategy. The telescope is moved using a right-going scan in azimuth, followed by a left-going scan, followed by a step in elevation. This pattern is repeated until a $\gtrsim$\,100 square degree patch of sky is observed. One such iteration is termed an \emph{observation}, whereas the patch of sky is termed a \emph{field}. 
In the only exception to this observing strategy, for two thirds of the data taken on one field (ra21hdec-50), the scan pattern instead consisted of scans in elevation at a series of fixed azimuth positions.
Each field was observed until a depth of $\lesssim$\,$18~\mathrm{\mu K}$-arcmin at 150~GHz was reached. Table~\ref{tab:fields} lists the 19 fields that comprise the full survey. For more information on the properties of the fields, see Table~1 of B15. 

Two-dimensional maps of the sky are made by binning the time ordered data (TOD), to which mild time-domain filtering has been applied, into 0.25-arcminute pixels.
This produces estimates of the 95 and 150~GHz sky, to which we apply a simultaneous spatial-spectral filter, yielding a filtered estimate of the tSZ sky, optimized for extracting cluster candidates. The candidates are identified using a peak-finding algorithm.
 
Due to the exact choice of field extent, as well as the finite footprint of the SPT-SZ array on the sky, the fields overlap slightly. 
The overlap regions only comprise $\sim2\%$ of the total survey area, and for simplicity we choose to treat the fields as fully independent. As a result of this treatment, we would expect to double-count roughly $2\%$ of cluster candidates, or seven to eight of our total number of candidates. In fact we find nine cases of candidates with a very high probability of having been detected in two different fields (i.e., 18~total candidates that correspond to nine physical clusters). There is no bias to our cosmological constraints from this treatment. Our uncertainties will be very slightly underestimated, but this has a negligible effect on our final cosmological constraints.
 
\subsubsection{SZ-Mass Scaling Relation Parameterization}
\label{sec:sr_sz}

We use the same functional form for the SZ-mass scaling relation as previous SPT cluster cosmology analyses 
(V10; \citealt{benson13}, hereafter B13; \citealt{reichardt13}, hereafter R13; \citealt{bocquet15}).

Briefly, we introduce two SZ parameters related to the cluster detection process. The first is the detection significance $\xi$. 
After generating filtered synthesized SZ maps at a series of filter scales, 
each map pixel value is divided by the RMS of the map in a strip that spans 90~arcminutes in declination. Then the maximum peak height over all filter scales is identified and 
defined as $\xi$, the SZ observable used in this paper.
Because this observable allows for a very well-understood selection function, and cosmological constraints are dominated by the unknown normalization of the observable-mass scaling relation, we find that this observable is preferred over going to a different SZ observable such as $Y_\mathrm{SZ}$.

The second parameter is the unbiased significance $\zeta$. It is defined as the value of $\xi$ that would be found in the absence of instrumental noise and astrophysical contaminants (including the SZ background). Due to the fact that $\xi$ is determined by maximizing the significance after searching in two-dimensional position space and source template size, the average $\xi$ found across many noise realizations is enhanced by those three degrees of freedom, resulting in the approximate relation
\begin{equation}
 \left< \xi \right>^2 = \zeta^2 + 3 .
\end{equation}
Since this relation cannot hold down to very low $\zeta$, we only model this maximization bias for $\zeta > 2$. We find that changing the location of this cutoff to $\zeta > 1.5$ or $\zeta > 2.5$ has negligible impact on the results presented in this work. Motivated by the definition of $\xi$, and the fact that the astrophysical contaminants and instrument noise are Gaussian to a high degree, we model $\xi$ as related to $\left< \xi \right>$ by unit-width Gaussian scatter. We then parameterize the \zetam\ scaling relation as
\begin{equation}
\label{eq:sr_sz}
 \zeta = A_\mathrm{SZ} \left( \frac{M}{3 \times 10^{14} M_{\odot} h^{-1}} \right)^{B_\mathrm{SZ}} \left(\frac{E(z)}{E(0.6)}\right)^{C_\mathrm{SZ}},
\end{equation}
with an additional log-normal intrinsic scatter parameter $\sigma_{\ln \zeta}$. The dimensionless Hubble parameter is denoted by $E(z)$. We follow B13 and R13 and apply Gaussian priors to these scaling relation parameters. However, the mean values and widths of the priors are updated to reflect results from the more recent cosmo-OWLS hydrodynamic simulations \citep{lebrun14}, which will be discussed in \S\ref{sec:sims}. Specifically, we use the Gaussian priors $A_\mathrm{SZ} = 5.38 \pm 1.61$, $B_\mathrm{SZ} = 1.340 \pm 0.268$, $C_\mathrm{SZ} = 0.49 \pm 0.49$, and $\sigma_{\ln \zeta} = 0.13 \pm 0.13$, while requiring $\sigma_{\ln \zeta} > 0.05$.
 
\subsection{By-Field Simulations}
\label{sec:sims}

Each of the 19 fields that comprises the 2500 square degree survey has slightly different properties, which we account for using two similar types of simulations. The first is to account for the fact that the map noise is slightly different in each of the fields. This causes the detection significance $\xi$ (the selection observable) of a cluster to relate to true underlying mass differently depending on the field in which the cluster was found. We model this by simulating the relation between true cluster mass and unbiased significance $\zeta$ separately for each field. The simulations contain several components.
\begin{enumerate}
  \item A Gaussian random field with a power spectrum equal to the CMB power spectrum calculated using the best-fit \lcdm\ parameters from \citet{keisler11}.
  \item A Gaussian random field meant to approximate the background of emissive point sources after the brightest sources are masked. The power spectrum is generated using three components. The first is a Poisson component modeling the radio point source population with an amplitude at $\ell=3000$ of $D_{3000}^r = 1.28 \mathrm{\mu K_{CMB}}^2$ at 150~GHz and spectral index $\alpha_r=-0.6$ (defined by flux $\propto \nu^\alpha$).  The second and third components model the Poisson and clustered dusty star-forming galaxy (DSFG) populations, respectively. The assumed spectral index is $\alpha_\mathrm{DSFG} = 3.6$ and the amplitudes are the best-fit values from \citet{shirokoff11}.
  \item Atmospheric and instrumental noise. These are simulated using the actual TOD. The coherent nature of the sky signal, modulated by the scan strategy, and the incoherent nature of the instrumental and atmospheric noise contributions allow us to estimate the map noise by simply subtracting the right-going scan TOD from the left-going scan TOD (up- and down-going in the case of most of the ra21hdec-50 field) in the map making process. This removes any coherent signal present on the sky. We furthermore apply a sign change to half the observations before they are coadded into the final map. By changing which signs are assigned, we generate different realizations of realistic noise maps.
  \item Maps of the Compton $y$ parameter from the cosmo-OWLS simulations in \citet{lebrun14}. We use maps from the AGN8.0 model and convert the Comptonization maps to CMB temperature units by integrating the measured frequency response of the instrument against a relativistic 5~keV tSZ spectrum.
\end{enumerate}
A modified version of the standard cluster-finding process is run on the sum of these four components, and the cluster candidates are recorded. In addition, the same spatial-spectral filter is applied to the maps only containing SZ signal. The latter is used to identify cluster candidates and define their SZ center. The amplitude in these filtered, SZ-only maps, divided by the standard deviation in a 90-arcminute strip in the filtered maps that include noise is precisely the unbiased significance $\zeta$ introduced in \S\ref{sec:sr_sz}. These candidates are then compared to the underlying halo catalog used to generate the mock SZ maps, which provide the cluster redshift. For each SZ-identified cluster candidate, we choose the nearest dark matter halo with $M_{200} > 5 \times 10^{13} \msun$.

With mass, unbiased significance and redshift in hand, we use the method of least absolute deviations to fit the \zetam\ relation from Equation~\ref{eq:sr_sz} over the range $z>0.25$, $M_{500} > 1 \times 10^{14} \msun$. 
We find nearly identical results when using a linear least squares method. 
The normalization parameter $A_\mathrm{SZ}$ differs from field to field by up to $30\%$, while the other scaling relation parameters only vary at the few percent level. We therefore model field variation with one parameter \fsf\ that renormalizes the overall scaling: $A_\mathrm{SZ} \rightarrow \fsf A_\mathrm{SZ}$. 
For consistency with previous publications, we normalize \fsf\, such that the weighted average over the three fields which share 100\% of the raw data used in R13 matches the weighted average in that work. We note that any other choice of normalization would simply alter the definition of $A_\mathrm{SZ}$, leaving all other results unaffected. The values of \fsf\, are shown in Table \ref{tab:fields}.

The other field-specific simulation is used to compute the expected false detection rate. At a detection threshold of $\xi>5$, approximately $5\%$ of cluster candidates are expected to be false. In order to simulate the rate of false detections, we perform the cluster-finding process on the simulated maps described above, omitting the SZ component. The resulting number of cluster candidates as a function of $\xi$ is recorded for 100 such simulations. To reduce shot noise from the finite number of simulated realizations, we model the false detection rate with the empirically chosen fitting function 
\begin{equation}
 \label{eq:false}
 N_\mathrm{false}(>\xi) = \falsenorm \exp \left ( -\falseslope(\xi-5) \right ) \ . 
\end{equation}
The values of \falsenorm\, and \falseslope\, are shown in Table \ref{tab:fields}. The total number of expected 
false detections per 2500~square degrees with $\xi > 5$ from these simulations is $18 \pm 4$, which is 
consistent with the number of optically unconfirmed cluster candidates, 21 (B15).

\begin{deluxetable}{ l c c c }
\tabletypesize{\scriptsize}
\tablecaption{The 19 SPT-SZ fields with the simulation-derived inputs to the cosmological analysis of the 2500~square degree survey \label{tab:fields}}
\tablewidth{200pt}
\tablehead{
\multicolumn{1}{l}{Name} &
\multicolumn{1}{c}{\falsenorm}  &
\multicolumn{1}{c}{\falseslope} &
\multicolumn{1}{c}{\fsf}
}
\startdata
\textsc{ra5h30dec-55}  &  16.79  &   4.60  &   1.33  \\
\textsc{ra23h30dec-55}  &  17.58  &   4.03  &   1.39  \\
\textsc{ra21hdec-60}  &  25.64  &   4.07  &   1.29  \\
\textsc{ra3h30dec-60}  &  20.53  &   4.70  &   1.25  \\
\textsc{ra21hdec-50}  &  25.28  &   4.14  &   1.11  \\
\textsc{ra4h10dec-50}  &  16.75  &   5.48  &   1.27  \\
\textsc{ra0h50dec-50}  &  20.76  &   5.11  &   1.14  \\
\textsc{ra2h30dec-50}  &  14.98  &   4.78  &   1.19  \\
\textsc{ra1hdec-60}  &  17.25  &   5.38  &   1.18  \\
\textsc{ra5h30dec-45}  &  15.91  &   4.81  &   1.08  \\
\textsc{ra6h30dec-55}  &  17.77  &   4.58  &   1.16  \\
\textsc{ra3h30dec-42.5}  &  16.85  &   4.31  &   1.20  \\
\textsc{ra23hdec-62.5}  &  14.90  &   4.92  &   1.18  \\
\textsc{ra21hdec-42.5}  &  17.11  &   4.49  &   1.15  \\
\textsc{ra1hdec-42.5}  &  18.41  &   5.55  &   1.19  \\
\textsc{ra22h30dec-55}  &  16.45  &   5.23  &   1.13  \\
\textsc{ra23hdec-45}  &  17.00  &   5.20  &   1.19  \\
\textsc{ra6h30dec-45}  &  14.78  &   4.23  &   1.16  \\
\textsc{ra6hdec-62.5}  &  16.53  &   4.70  &   1.18 
\enddata 
\tablecomments{The parameter \falsenorm\ describes the number of false detections expected above $\xi=5$ scaled to 2500 square degrees, while \falseslope\ describes the scaling of the number of false detections with $\xi$ as defined in Equation~\ref{eq:false}. \fsf\ describes the renormalization of $A_\mathrm{SZ}$ in the SZ-mass scaling relation of Equation~\ref{eq:sr_sz}. 
}
\end{deluxetable}
 
\section{Redshift Estimation}
\label{sec:redshifts}

As the SPT-SZ selection is essentially independent of redshift, we require optical and---for the highest-redshift systems---near-infrared (NIR) data to both confirm the SPT candidates as clusters and to obtain redshifts for these systems.
We provide a brief overview of this process here; for more details readers are referred to B15. 

As a first step, each SPT cluster candidate is visually inspected in imaging data from the Digitized Sky Survey (DSS),\footnote{http://archive.stsci.edu/dss/} as we have found most SPT clusters at redshift $z<0.5$ are visible in these scanned photographic plates.
These relatively low-redshift systems are then reimaged with 1 -- 2~m class telescopes to obtain robust confirmations and redshifts. 
Higher-redshift candidates not visible in the DSS (or designated non-confirmed after imaging on the small telescopes) are observed with 4 -- 6.5 m class telescopes.  
The latter observations are conducted in two passes:  first-pass observations  are designed to ensure $\ge 5 \sigma$ detection of  
$0.4$L$^*$ red-sequence galaxies 
(where L$^*$ is the characteristic luminosity that appears in the
\citet{schechter76} formulation of the galaxy luminosity function) 
at $z< 0.75$; higher-redshift clusters or candidates not confirmed in the first pass are also  observed (telescope resources permitting) in the optical and/or NIR to extend this redshift range to $z>0.9$.  
In total 69/78 of the highest-redshift ($z>0.75$) or non-confirmed systems received second-pass imaging;  the majority of these systems (58/78) were observed with  \spitzer /IRAC \citep[][PI: Brodwin]{fazio04} at 3.6~\um \ and 4.5~\um \  to a depth
sufficient for the $10 \sigma$ detection of  $0.4$L$^*$ galaxies at $z=1.5$ \citep{brodwin10}.

Following imaging, a few arcmin region around each SPT location is searched for an overdensity of red-sequence cluster galaxies. 
When such an overdensity is identified, we confirm the candidate  as a cluster and assign the system a redshift using a red-sequence model calibrated using the subset of SPT clusters with spectroscopic redshifts. 
For candidates not confirmed in our imaging data we compute a redshift ``lower limit''  corresponding to the highest redshift for which we would have detected the overdensity of red galaxies we require to confirm a cluster \citep{song12b}. 
The redshift range of the confirmed cosmological cluster sample is $0.25 \le z \lesssim 1.7$  with a median redshift $z_\textrm{med}=0.58$. 
Typical redshift uncertainties range from $\sigma_{z} \sim$0.02$\times(1+z)$ for the optical-based redshifts to $\sim$0.035$\times(1+z)$  for clusters with redshifts determined from {\it Spitzer}/IRAC observations. 

A large subset (31\%) of the SPT cosmological cluster sample has also been spectroscopically observed. 
Spectroscopic redshifts for 86~SPT clusters were obtained as part of a dedicated follow-up campaign using spectrographs on the Magellan telescope, the Gemini-South telescope, and the Very Large Telescope. 
We have also searched the literature for cluster counterparts and find an additional 21 clusters with reported spectroscopic redshifts. 
The spectroscopic sample spans almost the full redshift range of the cluster sample, from $0.26 < z \le 1.478$, with a median redshift of $z_\textrm{med}=0.53$.
The SPT spectroscopic follow-up effort is described in detail in \citet{ruel14}, and the redshifts are presented in B15.  

\section{\emph{Chandra} X-ray Dataset and Methods}
\label{sec:xray}
 
The X-ray data used in this work were originally presented in \citet{mcdonald13} and most were acquired as part of a \emph{Chandra} X-ray Visionary Project (PI: Benson). In general, exposure times were chosen to ensure 2000 X-ray counts, based on measured ROSAT fluxes (when available) or a combination of the $L_\mathrm{X}-M$ relation and mass estimates from the SZ signal. Data were obtained using the front-illuminated ACIS-I CCDs, and cleaned for background flares before applying calibration corrections using \textsc{ciao} v4.7 and \textsc{caldb} v4.6.8.

Global cluster properties (e.g., gas mass $M_{g,500}$ and X-ray temperature $kT_{500}$) for each cluster are derived in 
\citet{mcdonald13}, following closely the procedures described in \citet{andersson11}. For a detailed description of the X-ray analysis, the reader is directed to these works. In this work, we make use of the mass proxy \Yx, which is obtained from the measured gas mass and X-ray temperature. This temperature $kT_{500}$ is derived by first assuming some value of $\overdensityradius_{500}$ (i.e., 1~Mpc), and measuring the core-excised ($0.15~\overdensityradius_{500}$ to $1.0~\overdensityradius_{500}$) temperature within this radius. Using this temperature, we compute a new estimate of $\overdensityradius_{500}$ using the $T_\mathrm{X}-M$ relation from V09. The temperature is then measured within this radius, and the process is repeated until it converges.

We also measure the enclosed gas mass as a function of $\overdensityradius_{500}$, following the procedures described in \citet{mcdonald13b}. This involves measuring the X-ray surface brightness in the rest-frame energy range $0.7-2.0$~keV as a function of radius, and fitting a line-of-sight projected model for the electron density profile to these data, following \citet{vikhlinin06}. In converting from the electron density to the gas density, we assume $\rho_g = m_p n_e A/Z$, where $A = 1.397$ is the average nuclear charge and $Z = 1.199$ is the average nuclear mass. The enclosed gas mass within a given radius is simply the volume integral of the gas density profile out to the specified radius.
  
These two quantities are combined into the mass proxy \Yx. Since $\overdensityradius_{500}$ must be known to compute $M_{g,500}$, we numerically solve $\Yx=M_{g,500} kT_{500}$ together with the scaling relation given in Equation~\ref{eq:sr_x}. This numerical calculation is performed for each set of cosmological parameters that will be explored in \S\ref{sec:cosmo}.
 
\subsection{\yxm\ Scaling Relation Parameterization}
\label{sec:sr_x}
 
Following V09, B13, R13, and \citet{bocquet15}, we use \Yx\ as a proxy for the total cluster mass. We write the scaling relation as
\begin{equation}
 \label{eq:sr_x}
 \begin{split}
  \frac{M_{500}}{ 10^{14} M_{\odot} / h } = \left(A_\mathrm{X} h^{3/2} \left( \frac{h}{0.72} \right)^{\frac{5}{2} B_\mathrm{X} - \frac{3}{2}} \right) \\
  \times \left(  \frac{\Yx}{3 \times 10^{14} M_{\odot}\, \mathrm{keV}} \right)^{B_\mathrm{X}} E(z)^{C_\mathrm{X}},
 \end{split}
\end{equation}
where the parameters $A_\mathrm{X}$, $B_\mathrm{X}$, $C_\mathrm{X}$ describe the normalization, mass dependence and redshift dependence of the relation, respectively.\footnote{This equation differs slightly from B13 and R13 in order to more rigorously scale the V09 results if $B_\mathrm{X} \neq 0.6$ and $h \neq 0.72$ simultaneously. This modification has negligible effect on any of the results in this work.} As we did for the \zetam\ relation, we introduce a parameter $\sigma_{\ln \subYx}$ which models a log-normal intrinsic scatter in the \yxm\ relation. We allow the intrinsic scatter in the \yxm\ relation and the \zetam\ relation to be correlated, parameterizing this with the correlation coefficient $\rho_{\zeta,\subYx}$. We follow B13 and R13 and apply Gaussian priors of $B_\mathrm{X} = 0.57 \pm 0.03$, $C_\mathrm{X} = -0.40 \pm 0.20$, and $\sigma_{\ln \subYx} = 0.12 \pm 0.08$, respectively, as well as a uniform prior between $-0.98$ and 0.98 on $\rho_{\zeta,\subYx}$.  
  
\subsection{Prior on the Normalization of the \yxm\ Relation}
\label{sec:wlprior}
 
In B13 and R13, we used a prior on the normalization of the \yxm\ relation, $A_\mathrm{X}$, motivated from V09.  
In V09, the normalization was cross-checked against weak lensing mass estimates 
from \citet{hoekstra07} for the 10 clusters at $z < 0.3$ that, at the time, also had sufficient \chandra\ observations to measure \Yx.  In B13, 
this normalization was remeasured using more recent weak lensing mass estimates from \citet{hoekstra12}, and 
found to be consistent with the assumed calibration from V09. In this work, we revisit the \yxm\ normalization again, using the most recent 
weak lensing mass estimates from H15 to constrain any mean offset in the normalization of the original hydrostatic mass 
calibration of the \Yx-mass scaling relation from V09.  In addition, we also consider any systematic offset 
in the H15 weak lensing mass estimates, by comparing to alternative weak lensing mass estimates from WtG.

\begin{figure}
\begin{center}
 \includegraphics[angle=0,width=0.47\textwidth]{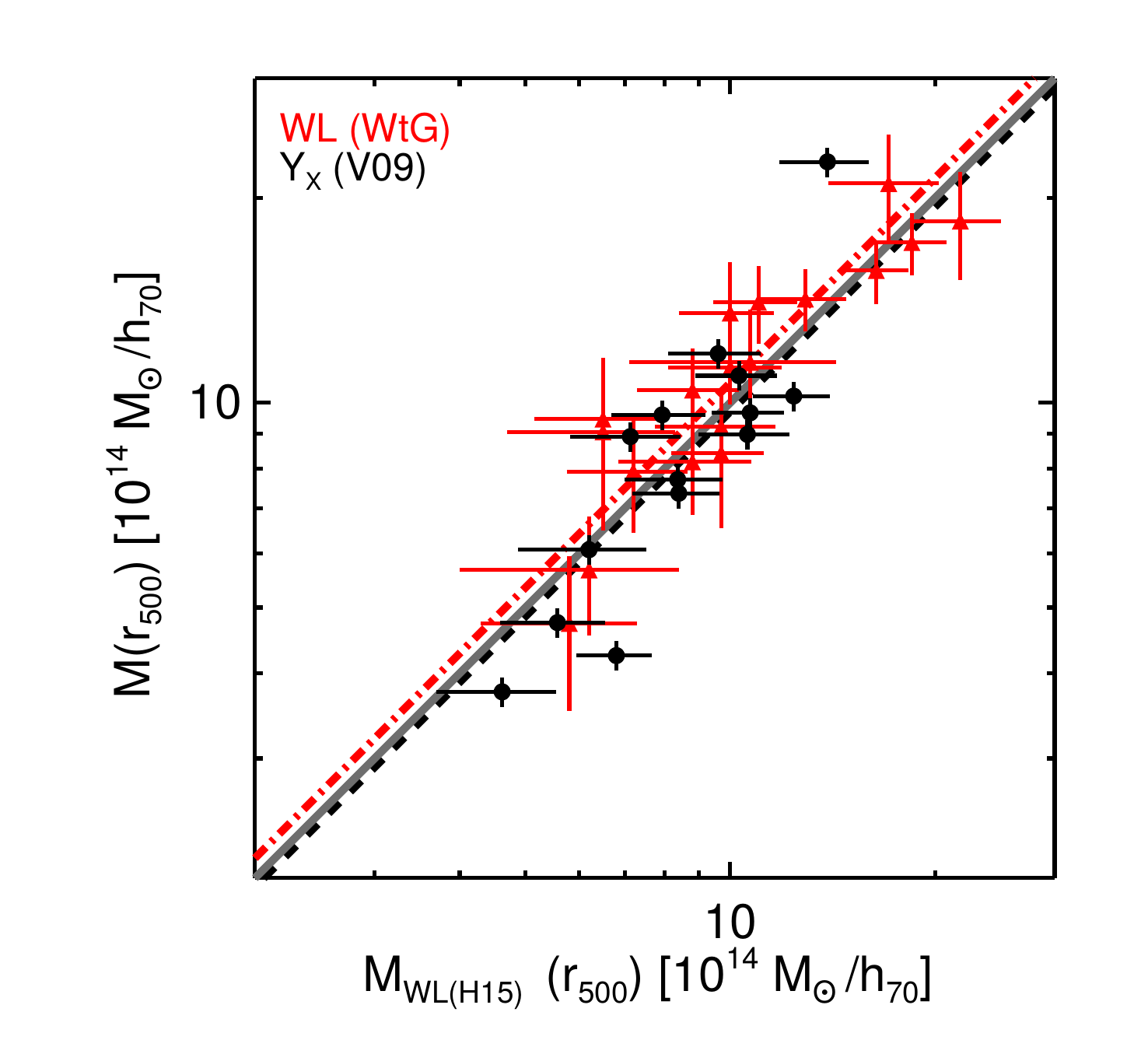}
\end{center}
\caption{A plot comparing cluster weak lensing and \Yx-based mass estimates.  
Plotted along the $y$-axis are weak lensing-based mass estimates from 
WtG (red) and \Yx-based mass estimates using the scaling from V09 (black).   
On the $x$-axis are weak lensing-based mass estimates from H15. For the H15/V09 comparison
(black points), we have re-estimated the H15 masses using the X-ray implied $\overdensityradius_{500}$.
The solid grey line shows a one-to-one relation, and the dashed black and red lines give the 
normalization implied for the bootstrap mean fit to the ratio of the masses (for details see text).}
\label{fig:yxm_norm}
\end{figure}

In Figure \ref{fig:yxm_norm}, we compare weak lensing-based mass estimates from H15 to \Yx-based mass estimates and to weak-lensing masses from WtG.
We have remeasured \Yx\ from archival \chandra\ X-ray data for the 14 clusters from H15 with sufficiently deep X-ray measurements, 
which we use to estimate a cluster mass ($M_{\rm \subYx}$) using the \Yx-mass scaling relation from V09.  For these clusters, we also remeasure the 
deprojected weak-lensing aperture mass from the 
H15 dataset using the X-ray implied $\overdensityradius_{500}$.   
Using the X-ray implied $\overdensityradius_{500}$ approximates 
what would happen if the weak lensing mass estimates were included in a joint \Yx - WL scaling relation fit because of the relatively low scatter of \Yx\ 
with cluster mass, and, 
as we will show later, 
the agreement in the resulting \yxm\ relation.
To compare H15 and WtG mass estimates at the same radius, we extract a WtG mass
for each of the 18~clusters in common at the value of $\overdensityradius_{500}$ 
implied by the H15 analysis. We specifically use the mass estimates computed 
by fitting an NFW model to the observed shear profiles from the WtG 
``color-cut'' method \citep{applegate14}.

In Table \ref{tab:mass}, we give three different measures of the relative cluster mass estimates between H15, WtG, and V09.  
First, we fit a scaling relation of the form $\ln(M_{\rm WL}) = A + \ln(M_{\rm \subYx})$ with a free log-normal intrinsic scatter, using the 
Bayesian linear regression fitting code from \citet{kelly07}.  
For comparison, we also estimate the bootstrap mean and median of the log of the mass ratio, which makes no assumption about the 
underlying scatter.  

First considering the mass ratio between H15 and V09, we find all three estimates imply a mass ratio near unity with the scaling relation, 
bootstrap mean, and bootstrap median giving ratios of $1.01 \pm 0.07$, $1.03 \pm 0.06$, and $1.12 \pm 0.05$, respectively.  This implies 
that the \Yx-based masses from V09 are in relatively good agreement with H15, with the H15 masses $\sim$1-12\% larger, 
depending on how the mass ratio is estimated.   
Next, we consider the mass ratio between WtG and H15.  We also find that all three estimates imply a mass ratio near unity with the scaling 
relation, bootstrap mean, and bootstrap median giving ratios of $1.06 \pm 0.07$, $1.07 \pm 0.07$, and $1.11 \pm 0.05$, respectively.  
Therefore the weak lensing-based mass estimates from H15 and WtG are also in relatively good agreement, with the WtG 
masses $\sim$6-11\% larger, comparable to the $7\%$ estimated systematic uncertainty from H15.  
The overlapping cluster sample between our V09-\Yx\ sample and WtG is smaller (eight clusters), and therefore 
has larger uncertainty on the mass ratio.  We find that the three different 
mass ratio estimates between WtG and V09 range from $1.15-1.24$.  This range is somewhat larger, but comparable with,
the expectation when 
considering the mass ratios of the larger samples used to calculate the WtG/H15 and H15/V09 mass ratios, which would predict 
a mass ratio between WtG and V09 of $\sim1.1-1.2$.

In summary, comparing to the original \yxm\ calibration from V09, we find that the weak lensing measurements imply a 
normalization that increases the \Yx-based cluster masses by a factor between 1.0 and 1.2, depending on the weak lensing analysis and dataset (H15, WtG), 
and the fit assumptions.  We therefore choose a Gaussian prior on the normalization of our \yxm\ relation of 
$A_\mathrm{X} = 6.35 \pm 0.61$, which increases the normalization from V09 by a factor of $1.1$ with an overall uncertainty 
of $10\%$.  We expect that this uncertainty includes the sum of our statistical and systematic uncertainty, because it brackets the 
range of values in the above comparison with the weak lensing measurements and is comparable to our naive expectation given the 
statistical uncertainty in the fit above and the systematic uncertainty estimated in H15.  This effectively gives us a purely 
weak lensing-derived mass scale, independent of the hydrostatic mass estimates. We expect our constraints on the mass scale to improve in future work using 
weak lensing observations of SPT-SZ survey clusters, which we will then be able to directly include in our cosmological analysis.  

Finally, we have also estimated the dependence of the weak-lensing and X-ray-derived mass estimates on cosmological parameters. Over the range of cosmological parameters explored, the ratio of the two varies negligibly compared to the uncertainty of $10\%$.

\begin{deluxetable}{ l c c c c}
\tabletypesize{\scriptsize}
\tablecaption{Mass ratios between weak lensing (H15, WtG) and \Yx\ (V09) based mass estimates.\label{tab:mass}}
\tablehead{
\multicolumn{1}{l}{Dataset} &
\multicolumn{1}{c}{$N_\mathrm{clust}$}  &
\multicolumn{1}{c}{Log-Normal} &
\multicolumn{1}{c}{Mean} &
\multicolumn{1}{c}{Median}
}
\startdata
H15/V09 & 14 & $1.01 \pm 0.07$ & $1.03 \pm 0.06$ & $1.12 \pm 0.05$ \\
WtG/H15 & 18 & $1.06 \pm 0.07$ & $1.07 \pm 0.07$ & $1.11 \pm 0.05$ \\
WtG/V09 & 8 & $1.15 \pm 0.10$ &  $1.17 \pm 0.13$ & $1.24 \pm 0.09$ 
\enddata
\end{deluxetable}

\section{Cosmological Analysis}
\label{sec:cosmo}
 
Armed with the abundance of clusters as a function of the observables $z$, $\xi$, and $\Yx$,
and models of the scaling of these observables with cluster mass, we are almost ready to place constraints on
cosmological parameters. For the final step, we need a theoretical framework for translating cosmological
parameters into a prediction for the number of clusters as a function of mass and redshift, and we need a 
statistical method for comparing those predictions to our observed abundance. For both of these requirements,
we closely follow the methods of V10, B13, and R13.

\subsection{Parameters to Cluster Abundance Predictions}
\label{sec:massfunction}

For a given set of cosmological parameters, we use the Code for Anisotropies in the Microwave Background
\citep[CAMB,][]{lewis00} to generate the matter power spectrum as a function of redshift. 
We use this power spectrum as input to the \cite{tinker08} halo mass function, with which we calculate the
number of dark matter halos as a function of spherical overdensity mass $M_{500}$ and redshift.

More recent results 
\citep[e.g.,][]{bhattacharya10, skillman14} confirm that, over the range of cluster masses and redshifts
considered in this work, the Tinker mass function is accurate to a level that is more than sufficient for 
the cosmological parameter constraints presented in this work.
For instance, even a shift as large as 10\% in the halo mass function at $M_{500} \sim 3 \times 10^{14}~\mathrm{\msun}$ 
would only affect the cosmological constraints presented in \S\ref{sec:lcdm} at the $\lesssim 0.2 \sigma$ level.

Given the Tinker mass function, we relate the
cluster mass to the observed quantities
$\xi$ and $\Yx$ using the scaling relations described in \S\ref{sec:sr_sz} and \S\ref{sec:sr_x}.

In our default cosmological analysis, we use the standard six-parameter \lcdm\ model, 
parameterized by the physical densities of baryons and cold dark matter at $z=0$, $\Omega_b h^2$
and $\Omega_c h^2$, the angular scale of the sound horizon at last scattering $\theta_s$, 
the tilt of the scalar power spectrum $n_s$, the amplitude of the scalar power spectrum 
$A_s$, and the optical depth to reionization $\tau$. We will often refer to values of 
parameters derived from combinations of these original six, such as $\sigma_8$, the amplitude 
of linear matter fluctuations on 8~Mpc$/h$ scales at $z=0$, \omm, the total matter density at $z=0$,
and $H_0 = 100h~\mathrm{km/s/Mpc}$, the value of the Hubble parameter at $z=0$. We explore three extensions
to \lcdm. We first explore a model in which the sum of the neutrino masses \summnu\ is a free 
parameter. In this work we use the prescription from \citet{constanzi13} to predict the 
effect of massive neutrinos on the halo mass function.
We also consider a cosmological model in which \summnu\ and \neff, the effective number of relativistic particle species,
are both free parameters, and a model in which $w$, the parameter describing the equation of state of dark energy, is allowed to vary.

\subsection{Cluster Likelihood Evaluation}
\label{sec:like}
 
To compare theoretical predictions for cluster abundance to the observations in this work, we closely follow the likelihood derivation of V10, B13, and R13. We differ from those analyses in this work by presenting
an efficient numerical technique that scales linearly rather than exponentially with the number of mass proxies.
 
As is appropriate for a cluster abundance measurement \citep{hu03a, holder06}, we start with the binned Poisson statistic.
For a given observable-space bin $x_i$, the probability of observing $n$ events, with expectation value $y(x_i)$ is
\begin{equation}
 P_i = \frac{e^{-y(x_i)}y(x_i)^n}{n!} .
\end{equation}
We choose to take the limit of small bins where $y$ becomes arbitrarily small at all $x_i$, and $n$ is zero except at the locations of $x_i$ where clusters have been observed \citep[][V09]{mantz08, mantz10a}. If we let $x_j$ denote the bin that contains the $j^\mathrm{th}$ cluster, we obtain
\begin{equation}
 \ln \mathcal{L} = \sum_i \ln P_i = - \sum_i y(x_i) + \sum_j \ln y(x_j) .
\end{equation}
In our case, we are dealing with the three-dimensional cluster observable space $\myvector{x}=\left[z, \xi, \Yx \right]$, though this method generalizes efficiently to a larger number of observables. Writing the model expectation value $y(x_i) = N(z_i, \xi_i, Y_{\mathrm{X},i})$ and going to one-dimensional indices gives
\begin{eqnarray}
 \label{eq:binned_like}
  \ln \mathcal{L} &=& - \sum_{i_1, i_2, i_3} N(z_{i_1}, \xi_{i_2}, Y_{\mathrm{X},{i_3}}) + \\
  \nonumber && \sum_j \ln N(z_j, \xi_j, Y_{\mathrm{X},j}) ,
\end{eqnarray}
where the $i_1, i_2, i_3$ sums again run over all possible values and $j$ only runs over the bins where clusters were detected. Going to the continuous limit would result in a divergent likelihood expression. This can be understood by the fact that the model is increasingly unlikely to produce our particular realization of the cluster catalog as we go to finer binning. The divergence can be removed by simply adding $- \ln \Delta z \Delta \xi \Delta \Yx$ to Equation \ref{eq:binned_like}. This quantity depends only on bin size, so that $\Delta \ln \mathcal{L}$ for different values of cosmological or scaling relation parameters remains meaningful. We then obtain the expression 
\begin{eqnarray}
 \label{eq:like}
  \ln \mathcal{L} &=& - \int dz d\xi d\Yx \frac{d N(z, \xi, Y_{\mathrm{X}})}{dz d\xi d\Yx} + \\
  \nonumber && \sum_j \ln \frac{d N(z_j, \xi_j, Y_{\mathrm{X},j})}{dz d\xi d\Yx} .
\end{eqnarray}
V10, B13, and R13 evaluated this expression on either a two-dimensional or three-dimensional uniformly spaced grid. Each dimension was gridded into several hundred points, such that the computational cost rises exponentially with the number of mass proxies. The computational cost was trivial for V10, challenging for B13 and R13, and likely untractable in future work if another mass proxy 
is added. The techniques used in \citet{mantz10a} and \citet{vikhlinin09} also scale exponentially with the number of mass proxies.
 
Instead of computing Equation \ref{eq:like} on a uniformly spaced grid, we use the following numerical techniques. First, we note that the follow-up mass proxy $\Yx$ immediately integrates out of the first term. This applies to any mass proxy that does not affect the survey selection function. Therefore, we only need to perform a two-dimensional integral
\begin{equation}
  - \int dz d\xi d\Yx \frac{dN(z, \xi, Y_{\mathrm{X}})}{dz d\xi d\Yx} 
  = - \int_{z_\mathrm{cut}}^\infty dz \int_{\xi_\mathrm{cut}}^\infty d\xi \frac{dN}{dz d\xi} .
\end{equation}
We evaluate this expression with
\begin{eqnarray}
  \int_{z_\mathrm{cut}}^\infty dz \int_{\xi_\mathrm{cut}}^\infty d\xi \frac{dN}{dz d\xi} \hspace{4cm} \\
  \nonumber = \int_{z_\mathrm{cut}}^\infty dz \int_{0}^\infty dM \frac{dN}{dz dM} P(\xi > \xi_\mathrm{cut} | M) , 
 \end{eqnarray}
where $P(\xi > \xi_\mathrm{cut} | M)$ is simply the significance cut modelled through the scaling relation as described in  \S\ref{sec:sr_sz}.

The second term of Equation \ref{eq:like} is more challenging to evaluate. Employing a more compact notation with $\myvector{x}=\left[z, \xi, \Yx \right]$, we start by writing the expectation density as an integral over the mass function, 
\begin{eqnarray}
 \label{eq:clustint}
  \frac{dN(\myvector{x}_j)}{d\myvector{x}} &=& \int d\myvector{x}' \int d\ln M' \\
  \nonumber && \prod_i P(x_{i,j}|x_i') P(\myvector{x}' | z, \ln M') \frac{dN}{dz d\ln M'} .
\end{eqnarray}
The probability density functions $P(x_{i,j}|x_i')$ describe the measurement error of the $i^\mathrm{th}$ mass proxy (with the measurements being independent). The $P(\myvector{x}' | z, \ln M')$ describe the joint scaling relations, implemented with multi-dimensional log-normal intrinsic scatter. In our case of two mass proxies ($\xi$, \Yx), this is implemented with three parameters: two parameters describing the marginal variances and one correlation coefficient, each of which is marginalized over in this work.
 
The computational bottleneck for evaluating the cluster likelihood in the presence of several mass proxies lies in evaluating Equation \ref{eq:clustint}. Considering a slice in redshift, we have to perform an ($N_\mathrm{obs}+1$)-dimensional integral. To do so, we perform Monte-Carlo integration, randomly sampling the mass function $\frac{dN}{d\ln M'}$ with points drawn from the probability distribution $\prod_i P(x_{i,j}|x_i') P(\myvector{x}' | \ln M')$. Let $\ln \myvector{m}'$ denote the integration variables corresponding to the nominal mass estimates (i.e. $\myvector{x}'$ substituted into the observable-mass scaling relations). We can then write
\begin{eqnarray}
 \label{eq:inttrick_start}
 P(\ln \myvector{m}' | \ln M') = \frac{1}{\sqrt{(2 \pi)^{N_\mathrm{obs}} |\covm|}} \hspace{3cm} \\
 \nonumber \exp \left( - \frac{1}{2} \left( \ln \myvector{m}' - \ln M' \right)^\top \covm^{-1} \left( \ln \myvector{m}' - \ln M' \right) \right) ,
\end{eqnarray}
where $\covm$ is the mass proxy covariance matrix containing the intrinsic scatter and correlation coefficient parameters introduced in \S\ref{sec:sr_sz} and \S\ref{sec:sr_x} such that $\covm_{kl} = \left\langle (\ln m_k' - \ln M') (\ln m_l' - \ln M') \right\rangle$. 
 
Now, we wish to obtain samples of the integration variable $\ln M'$, given the location at which we are attempting to evaluate $dN(\myvector{x}_j)/d\myvector{x}$. When the probabilities $P(x_{i,j}|x_i')$ involved are Gaussian or log-normal, they are constructed to be correctly normalized. Due to the treatment of maximization bias in \S\ref{sec:sr_sz}, the normalization condition needs to be explicitly applied to the SZ scaling relation. This yields a factor of 
\begin{equation}
 \int \exp \left( - \frac{1}{2} \left( \langle \xi \rangle - \zeta \right)^2 \right) \frac{1}{\zeta} d\zeta ,
\end{equation}
which we evaluate numerically using the trapezoid rule. In order to efficiently draw samples we first, for each $i$, draw samples from the measurement error $P(x_{i,j}|x_i')$, which is assumed to be independent for each $i$. We then substitute these values into the scaling relations to obtain an ensemble of $\ln \myvector{m}'$. The remaining task is then to draw random deviates $\ln M'$ that follow the probability distribution explicitly shown in Equation \ref{eq:inttrick_start}, given each value of $\ln \myvector{m}'$. To do so, we note that
\begin{equation}
 \label{eq:inttrick}
 \begin{array}{l}
 \left( \ln \myvector{m}' - \ln M' \right)^\top \covm^{-1} \left( \ln \myvector{m}' - \ln M' \right) \\
 = \Sigma_{ij} (\ln m'_i - \ln M') (\ln m'_j - \ln M') (\covm^{-1})_{ij} \\ 
 = \Sigma_{ij} \ln m'_i \ln m'_j (\covm^{-1})_{ij} \\ - 2 \ln M' \Sigma_i \ln m'_i \Sigma_j (\covm^{-1})_{ij} + \ln M'^2 \Sigma_{ij} (\covm^{-1})_{ij} \\
 = T_0 - 2 T_1 \ln M' + T_2 \ln M'^2  \\
 = \left( \frac{\ln M' - T_1/T_2}{1/\sqrt{T_2}} \right)^2 + T_0 - \frac{T_1^2}{T_2} , \\
 \end{array}
\end{equation}
where $T_0 = \Sigma_{ij} \ln m'_i \ln m'_j (\covm^{-1})_{ij}$, $T_1 = \Sigma_i \ln{m'_i} \Sigma_j (\covm^{-1})_{ij}$, and $T_2 = \Sigma_{ij}  (\covm^{-1})_{ij}$,
which is quadratic in $\ln M'$, such that Equation \ref{eq:inttrick_start} is a log-normal distribution in $M'$ with a known mean, width and normalization. We compute $T_0$, $T_1$, and $T_2$ explicitly and sample from the resulting distribution.
 
Having obtained samples of $\ln M'$, which we denote as $\ln \tilde{M'}$, we average the mass function 
\begin{equation}
 \frac{1}{N_\mathrm{samples}} \sum_{k} \left. \frac{dN}{d \ln M'} \right|_{\ln \tilde{M'}_k} \sqrt{\frac{T_2}{2 \pi}} \frac{\exp{-\frac{1}{2}\left(T_0 - \frac{T_1^2}{T_2} \right)}}{\sqrt{(2 \pi)^{N_\mathrm{obs}} |\covm| }}
\end{equation}
and combine with Equation~\ref{eq:inttrick} to obtain an estimate for Equation~\ref{eq:clustint}. Our implementation of this estimator has been demonstrated to be unbiased through extensive simulations, with well-behaved residuals. The error on the mean is found to decrease as the inverse square root of the number of deviates drawn. In practice, for two mass proxies and \nclust\, cluster candidates, we draw $10^4$ deviates per cluster candidate, resulting in RMS noise on the likelihood surface of $\Delta (-2\ln \mathcal{L}) \lesssim 0.1$ near the maximum likelihood. For the cluster catalog used in this work, the execution time is approximately one second on a single CPU thread. 
This likelihood module has been checked against an independent implementation, based on the estimator presented in \citet{bocquet15}. The comparison shows agreement to well within the 1-$\sigma$ uncertainties on the cosmological parameter constraints found in this work. 
 
\subsection{External Datasets}
\label{sec:external}

In \S\ref{sec:constraints}, we will discuss the cosmological constraints obtained using the analysis laid out so far. We will evaluate the compatibility of the cluster data with other datasets, as well as show the improvements in parameter constraints when the cluster dataset is combined with other datasets. The primary external data we use are CMB power spectrum measurements, measurements of BAO from galaxy surveys, and distances to Type Ia supernovae (SNe).

The canonical CMB power spectrum data we use is the temperature-temperature power spectrum from the \planck\, 2013 release, combined with low-$\ell$ polarization information from \wmap\, \citep[][hereafter \planck+WP]{planck13-16}. The qualitative results from this work will be very similar to those that would have been obtained by considering the \planck\, 2015 dataset, since the constraints on $\sigma_8$ and $\omm$ from the temperature and low-$\ell$ polarization power spectrum are very similar between the two \planck\, releases. In \S\ref{sec:lcdm}, we will briefly review the effect of substituting the best CMB power spectrum data from before the \planck\, 2013 release, using instead the combination of \wmap9 \citep{hinshaw13} and SPT \citep[][hereafter S13]{story13} CMB power spectrum data. 

We sometimes use a prior on the angular scale of the sound horizon $\theta_s$. This is a powerful piece of cosmological information, and relies only very weakly on the details of the CMB analysis, since it is sensitive to peak positions. We use the \planck+WP measurement, though we conservatively increase the uncertainty by a factor of five to $100\theta_\mathrm{MC} = 1.0413 \pm 0.0034$.\footnote{$\theta_\mathrm{MC}$ is the approximation to $\theta_s$ used in CosmoMC \citep{lewis02b}.} Since $\theta_s$ is so well-measured, none of the results presented in this work that use a prior on $\theta_s$ are sensitive to the specific value of the assumed uncertainty.

When BAO data are used in this work, we use the SDSS-III BOSS results from data release 11. Specifically, we use the measurements of the parameter combination $D_V / r_s$ at $z=0.32$ (LOWZ) and $z=0.57$ (CMASS) from \citet{anderson13}.

When considering cosmologies with a free parameter describing the dark energy equation of state, we compare to and contrast with the SN results from the joint likelihood analysis of the SDSS-II and SNLS SN samples from \citet{betoule14}.
 
Cluster abundances do weakly constrain the baryon density, but for our main results we choose to
adopt a prior based on big-bang nucleosynthesis calculations and deuterium abundance measurements
of $\Omega_b h^2 = 0.02202 \pm 0.00045$ \citep{cooke14}. This uncertainty is subdominant for all 
results presented in this work.

Finally, where stated, we adopt a prior of $H_0 = 73.8 \pm 2.4~\mathrm{km/s/Mpc}$ from the direct $H_0$ measurements of \citet{riess11}. We find that changing this prior by $1\sigma$ affects the resulting $\sigma_8$ constraint by $\lesssim0.2\sigma$, while the \omm\ constraint is affected by $\sim0.4\sigma$. These two effects are anti-correlated such that the commonly used parameter combination \seight\ is highly insensitive to the assumed value of $H_0$.
 
 \section{Cosmological Constraints}
 \label{sec:constraints}
 
In this section, we discuss the parameter constraints obtained using the data and methods described in the 
previous sections. We first explain which parameters and combinations of parameters are most strongly 
constrained by cluster abundance measurements. We then discuss the constraints from this cluster sample, 
both in comparison to and in combination with other cosmological datasets, for different choices of 
cosmological models. 
We assume spatial flatness throughout this work.

\subsection{Parameter Sensitivity to Cluster Abundance}
\label{sec:paramsens}

A measurement of cluster abundance as a function of redshift provides constraints on cosmological parameters through several mechanisms. First, the total number of clusters found strongly depends on the matter density and the amplitude of the matter power spectrum \citep[e.g.,][]{bahcall92,white93b}. As shown in, e.g., \citet{white93b}, a simple 
spherical-collapse model for the halo mass function \citep{press74} indicates that the total number of
clusters should most strongly constrain the parameter combination $\sigma_8 \omm^\alpha$, 
where $\alpha$ is related to the local slope of the matter power spectrum at the mean mass of the cluster catalog.
This prediction has been empirically borne out in many cluster abundance studies, including this one.
 
The redshift dependence of the cluster abundance contains information on the growth function, as well as a dependence on the cosmic volume surveyed. 
This combination of sensitivity to growth and volume provides a unique constraint on parameters that 
affect the expansion history, notably $w$, the equation of state of dark energy \citep[e.g.,][]{haiman01}.
In the case of $w$, the effect of changing this parameter on the cluster abundance actually changes 
sign at $z \sim 1$ \citep[e.g., Fig.~1 of][]{mohr05}.
The \nclust\, cluster candidates presented in this work provide a large enough sample to constrain cosmological parameters by measuring the evolution of cluster abundance. However, 
this constraint is also potentially limited by knowledge of the observable-mass scaling relations, 
in this case their evolution with redshift.
This is further discussed in the context of dark energy in \S\ref{sec:wcdm}.
 
In addition to their effect on the true abundance of clusters as a function of mass and redshift, cosmological parameters 
also affect the cluster observables through their influence on the SZ and X-ray scaling relations.
Specifically, while the mass estimate inferred from the \zetam\ relation depends weakly on redshift,
the mass estimate inferred from the \yxm\ scaling relation depends strongly on the angular diameter distance to the cluster.
   
\subsection{$\Lambda$CDM}
\label{sec:lcdm}
 
\begin{figure}
\begin{center}
 \includegraphics[angle=0,width=0.47\textwidth]{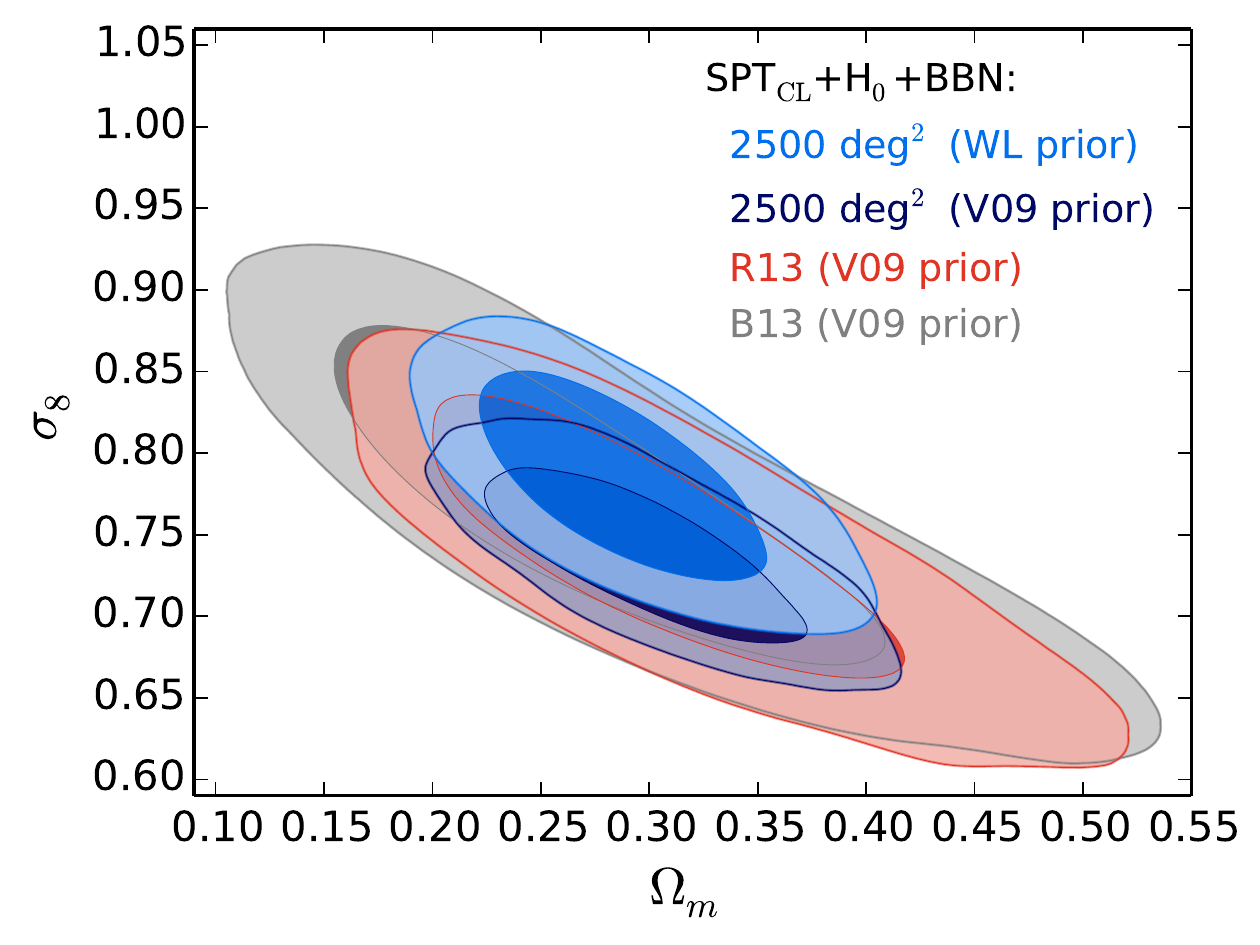}
\end{center}
\caption{Comparison of cluster constraints on $\sigma_8$ and \omm\, from this work with those from previous SPT publications. The B13 analysis (outermost, gray contours) used 18~clusters, 14~of which have \chandra\, observations. The number of clusters increased to 100~in R13 (red contours), whereas this work uses \nclust\, cluster candidates, \nclustx\, of which have high-quality \chandra\, observations. If we adopt the same observable-mass priors as B13 and R13, we obtain the innermost, purple contours. However, the main results in this paper assume a new weak lensing-based prior on the 
X-ray scaling relation normalization, which changes the central value by 10\% and increases the 1$\sigma$ uncertainty slightly from 9\% to 11\%. The $\sigma_8$-\omm\, constraints using this prior and the current cluster data
are shown by the light-blue contours.}
\label{fig:b13_r13_2500d_14}
\end{figure}

In this section, we present constraints on the parameters of the \lcdm\ model. 
Because not all parameters of this model are well constrained by cluster counts alone, we adopt 
priors on some of the six parameters ($\Omega_b h^2$, $\Omega_c h^2$, $\theta_s$, $n_s$, 
$A_s$, and $\tau$, defined in \S\ref{sec:massfunction}).
The cluster likelihood is insensitive to the optical depth to reionization $\tau$ as well as to the primordial scalar spectrum power law index $n_s$, once an appropriate pivot point is chosen. Therefore, when discussing cluster constraints without the inclusion of CMB temperature power spectrum data, we fix these parameters to the best-fit values from \citet{planck13-16}.
 
\begin{figure*}
\begin{center}
 \includegraphics[angle=0,width=0.97\textwidth]{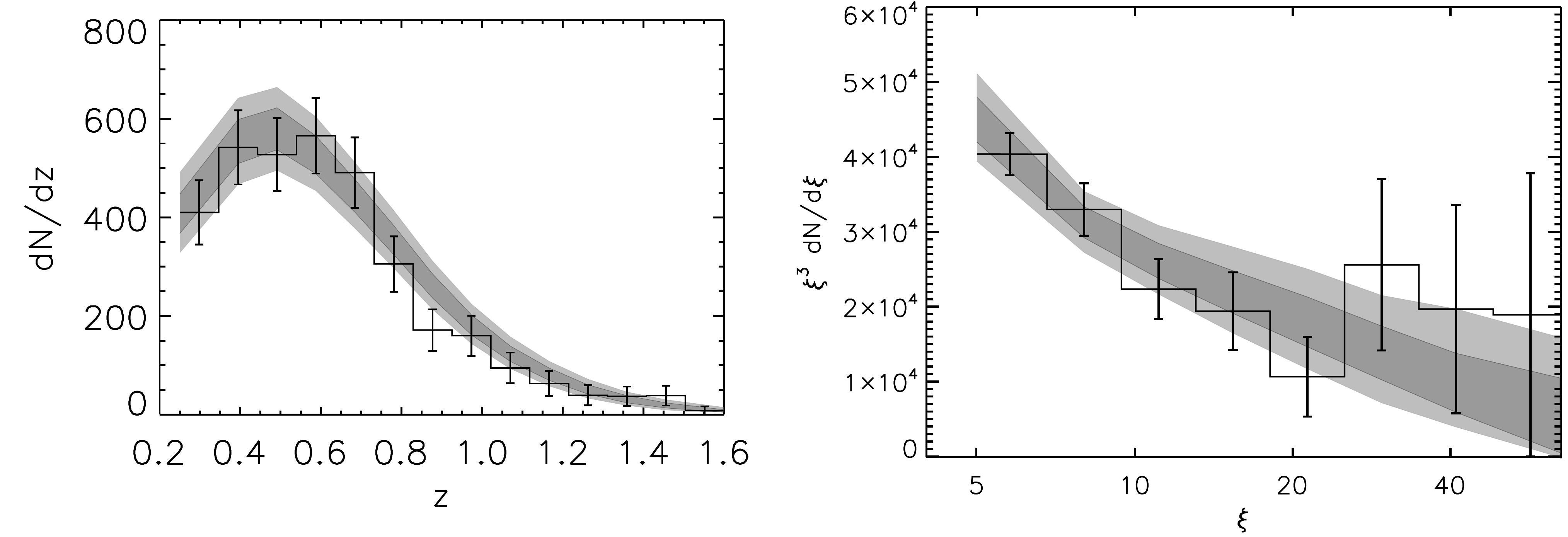}
\end{center}
\caption{Number density of clusters as a function of redshift (left panel) and of the SPT-SZ mass proxy $\xi$ (right panel). The data points show the measured abundance with $\sqrt{N}$ error bars. The grey bands show the $68\%$ and $95\%$ allowed model ranges after marginalizing over all cosmological and scaling relation parameters in the \lcdm\ model with the \sptcl+$H_0$+BBN dataset. In the right hand panel, the $\xi$ axis is shown on a logarithmic scale and the abundance axis has been multiplied by three powers of $\xi$ in order to visualize the abundance over a range of $\xi$ values despite the extreme steepness of the measured mass function.}
\label{fig:dndX_twocol}
\end{figure*}
  
These two prior constraints, in addition to the $H_0$ and BBN priors discussed in \ref{sec:external},
leave us two remaining degrees of freedom in the \lcdm\ model, which
we choose to express as \omm\, and $\sigma_8$ (though in exploring the likelihood surface we actually vary the base parameters $\Omega_b h^2$, $\Omega_c h^2$, $\ln 10^{10} A_s$ and 
$H_0$). 
We fix the species-summed neutrino mass \summnu\, to $0.06$~eV, the minimum allowed value from terrestrial measurements of squared neutrino mass differences \citep[see e.g.][for a review]{gonzalezgarcia12}.
Figure \ref{fig:b13_r13_2500d_14} shows the constraints on $\sigma_8$ and \omm\, from B13, R13 and this work. The large increase in the number of clusters primarily reduces the uncertainty on \omm\, by constraining the shape of the halo mass function. The uncertainty on the parameter combination \seight\ is set by the knowledge of the overall mass scale, rather than the number of clusters, since Poisson errors on this number are subdominant. The improvement in the \seight constraint results from the increase in the number of clusters with \chandra\ follow-up data. These two effects result in tighter 
joint constraints on the two parameters than we obtained in previous cluster analyses, if we use the same prior 
on the observable-mass relations. 
This can be seen in Figure \ref{fig:b13_r13_2500d_14}: the innermost, purple contours use the
current cluster sample and the same observable-mass priors as B13 and R13.
However, in contrast to B13 and R13 we choose to use a weak lensing-based prior on the overall mass scale for our baseline results, as discussed in \ref{sec:wlprior}. This results in slightly
degraded constraints, in particular on $\sigma_8$, as can be seen from the light-blue contours in 
Figure \ref{fig:b13_r13_2500d_14}. With the current cluster sample, the stated priors on cosmological 
parameters, and the updated priors on the observable-mass relations, 
we obtain 
\begin{equation}
 \sigma_8 = \constrainthsptclsigmaeight ,
\end{equation}
\begin{equation}
 \omm = \constrainthsptclomegam ,
\end{equation}
and
\begin{equation}
 \seight = \constrainthsptclseight.
\end{equation}
The \lcdm\ parameter constraints, including scaling relation parameters, are shown in Table~\ref{tab:lcdmresults}.

\begin{deluxetable*}{ l c c c c }
\tabletypesize{\scriptsize}
\tablecaption{Constraints on cosmological and scaling relation parameters assuming a \lcdm\ cosmology \label{tab:lcdmresults}}
\tablewidth{0pt}
\tablehead{
\multicolumn{1}{l}{Parameter} &
\multicolumn{1}{c}{Prior} &
\multicolumn{1}{c}{\sptcl+$H_0$+BBN}  &
\multicolumn{1}{c}{\sptcl+BAO+BBN} &
\multicolumn{1}{c}{\sptcl+\planck+WP+BAO}
}
\startdata

$A_\mathrm{SZ}$ & $5.38 \pm 1.61$ & $ 4.842 \pm  0.913$ & $ 4.936 \pm  0.955$ & $ 3.531 \pm  0.273$ \\
$B_\mathrm{SZ}$ & $1.340 \pm 0.268$ & $ 1.668 \pm  0.083$ & $ 1.660 \pm  0.064$ & $ 1.661 \pm  0.060$ \\
$C_\mathrm{SZ}$ & $0.49 \pm 0.49$ & $ 0.550 \pm  0.315$ & $ 0.864 \pm  0.159$ & $ 0.733 \pm  0.123$ \\
$\sigma_{\ln \zeta}$ & $0.13 \pm 0.13$ & $ 0.199 \pm  0.069$ & $ 0.201 \pm  0.070$ & $ 0.203 \pm  0.066$ \\
$A_\mathrm{X}$ & $6.38 \pm 0.61$ & $ 6.235 \pm  0.514$ & $ 6.316 \pm  0.505$ & $ 7.030 \pm  0.341$ \\
$B_\mathrm{X}$ & $0.57 \pm 0.03$ & $ 0.491 \pm  0.023$ & $ 0.493 \pm  0.023$ & $ 0.498 \pm  0.021$ \\
$C_\mathrm{X}$ & $-0.40 \pm 0.20$ & $-0.251 \pm  0.127$ & $-0.280 \pm  0.122$ & $-0.174 \pm  0.102$ \\
$\sigma_{\ln \subYx}$ & $0.12 \pm 0.08$ & $ 0.162 \pm  0.070$ & $ 0.160 \pm  0.069$ & $ 0.154 \pm  0.064$ \\
$\rho_{\zeta, \subYx}$ & $[-0.98,0.98]$ & $-0.147 \pm  0.458$ & $-0.136 \pm  0.465$ & $-0.204 \pm  0.443$ \\
$\omm$ &  & $ 0.289 \pm  0.042$ & $ 0.306 \pm  0.010$ & $ 0.304 \pm  0.007$ \\
$\sigma_8$ &  & $ 0.784 \pm  0.039$ & $ 0.768 \pm  0.030$ & $ 0.820 \pm  0.009$ \\
$\seight $ &  & $ 0.797 \pm  0.031$ & $ 0.797 \pm  0.030$ & $ 0.850 \pm  0.013$ 
\enddata 
\end{deluxetable*} 
 
In the left- and right-hand panels of Figure~\ref{fig:dndX_twocol}
we show the cluster abundance as a function of redshift and detection significance, respectively. Both show one-dimensional representations of the observable-space mass function. The data points, with approximate ($\sqrt{N}$) error bars shown, are independent of cosmological and scaling relation parameters. The points are independent between bins for $dN/d\xi$, and nearly independent for $dN/dz$, where clusters without spectroscopic redshift information can contribute to multiple bins. 
 
Shifts in the parameters $\sigma_8$ and $A_\mathrm{SZ}$ result in roughly global shifts in the amplitude of both curves, simultaneously. The scaling relation parameter $B_\mathrm{SZ}$ induces a roughly power-law tilt in $dN/d\xi$, and parameters such as $C_\mathrm{SZ}$ and \omm\, induce tilts in $dN/dz$. This visualization shows two important ways in which the model, marginalized over a large number of scaling relation and cosmological parameters, is tested for agreement with the data.
 
\begin{figure}
\begin{center}
 \includegraphics[angle=0,width=0.47\textwidth]{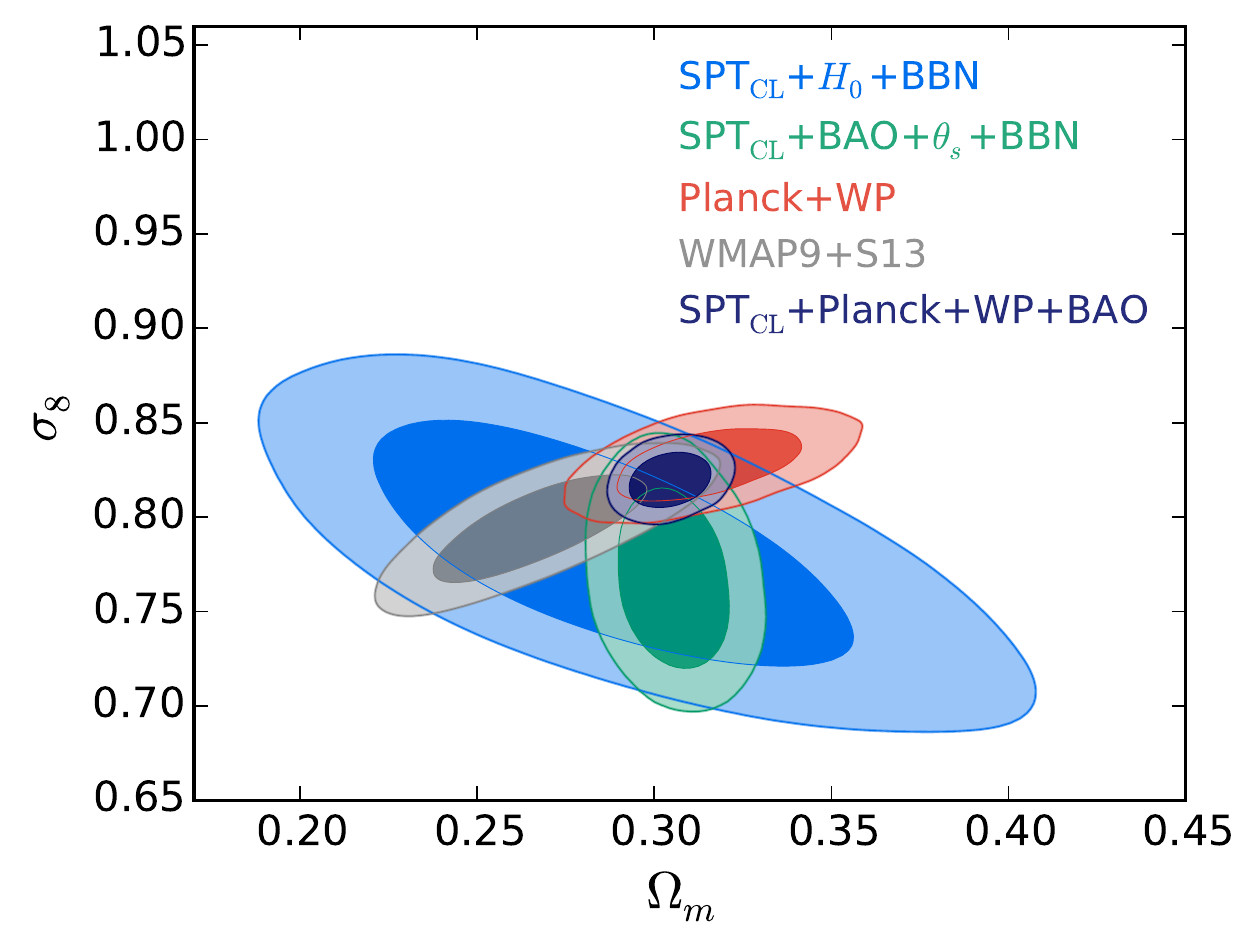}
\end{center}
\caption{Comparison of cluster constraints on $\sigma_8$ and \omm\, to constraints from primary CMB anisotropies, assuming a \lcdm\ cosmology. The cluster constraints, when combined with either the $H_0$ or BAO$+\theta_s$ prior, are in agreement with the CMB datasets.}
\label{fig:2500d_cmb_comparison}
\end{figure}
 
Figure \ref{fig:2500d_cmb_comparison} shows the cluster constraints on the \lcdm\ model in
the $\sigma_8$-\omm\, plane, when combined with either the $H_0$+BBN prior or the
BAO$+\theta_s$+BBN prior. We also show the constraints from CMB power spectrum measurements from 
\planck+WP and 
\wmap9+S13 data. 
The 68\% confidence regions from the cluster constraints and the CMB power spectrum constraints overlap. We proceed to adopt the \planck+WP dataset as the baseline CMB dataset for the remainder of this work. 
We also combine the \sptcl+\planck+WP+BAO datasets to obtain joint \lcdm\ parameter constraints.
Finally, we note that the $\sigma_8$ constraint obtained using the \sptcl\ data in Table \ref{tab:lcdmresults} 
is within $1 \sigma$ of the value recently reported by the \planck\ collaboration for the full-mission data (their 2015 TT+lowP+lensing dataset) of $\sigma_8 = 0.815 \pm 0.009$ .
 
\subsubsection{Constraints on Scaling Relation Parameters}
\label{sec:sr_params}
 
\begin{figure*}
\begin{center}
 \includegraphics[angle=0,width=0.97\textwidth]{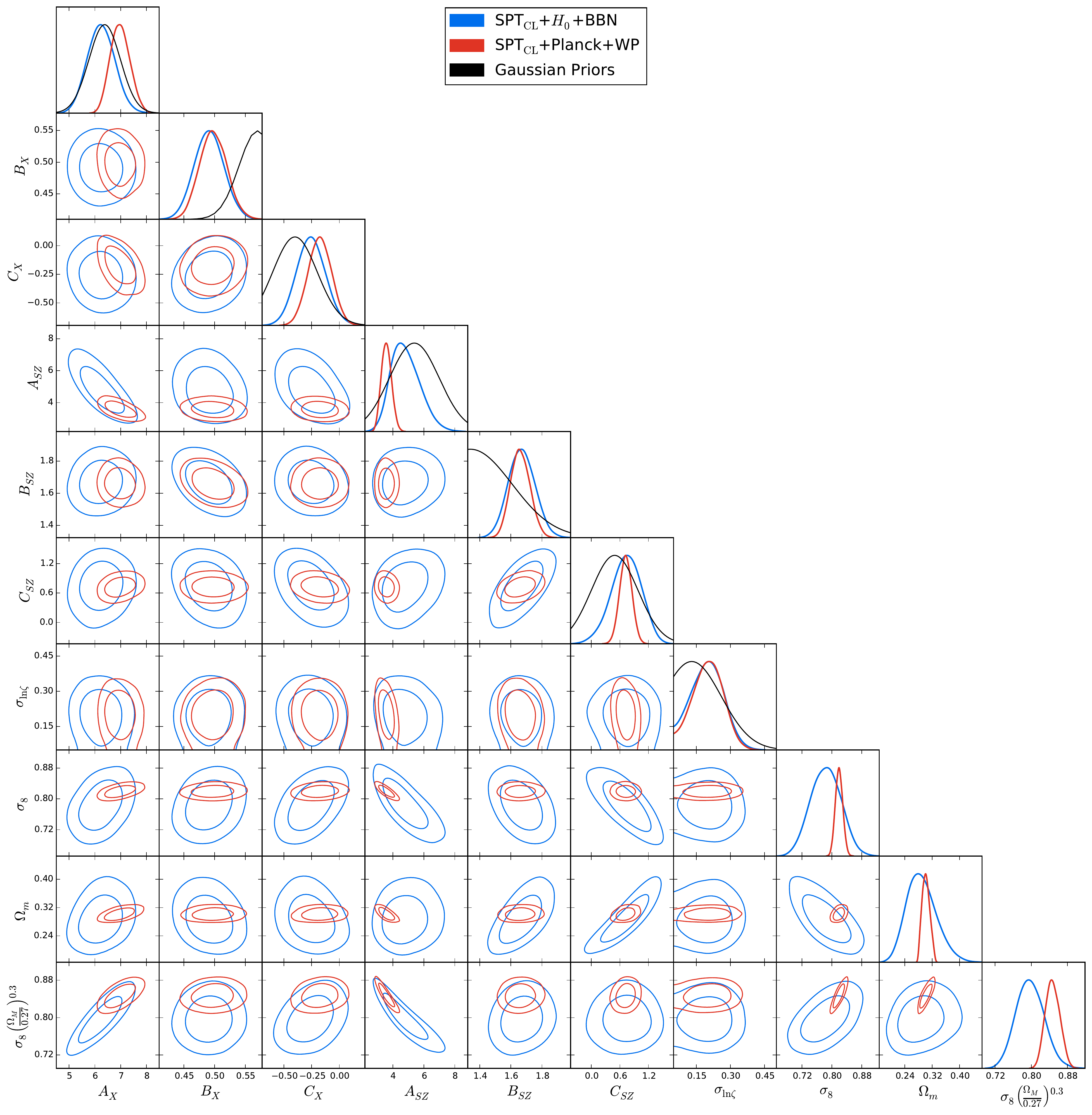}
\end{center}
\caption{Contour triangle plot showing the degeneracies between scaling relation parameters and cosmological parameters. Parameters $\Omega_b h^2$, $H_0$, $n_s$, $\sigma_{\ln \subYx}$, and $\rho_{\zeta,\subYx}$ are marginalized over and not shown since they are primarily constrained by priors, or by the \planck\ data. The cluster likelihood is nearly flat over the explored range of these parameters.}
\label{fig:sr_sptcl_planck}
\end{figure*}

While the main focus of this work is on the cosmological constraints, the nuisance parameters are of interest themselves, both in terms of the cluster scaling relation parameter constraints and their degeneracy with the cosmological parameters. In Figure \ref{fig:sr_sptcl_planck}, we show the marginalized posterior in the multi-dimensional parameter space.
 
The dominant systematic uncertainty limiting the cosmological constraints from the SPT-SZ cluster sample is the uncertainty of the overall cluster mass scale.  This can be seen as a strong degeneracy between $A_\mathrm{X}$ and $\seight$, which are $85\%$ correlated given the \sptcl+$H_0$+BBN dataset.  
The second-most important source of systematic uncertainty in the cosmological constraints is from the parameter characterizing the redshift evolution of the scaling relation, $C_\mathrm{SZ}$. In a \lcdm\ cosmology, it is highly degenerate with \omm, correlated at 87\% when considering the \sptcl+$H_0$+BBN dataset.

The parameters that shift most significantly away from their priors are the $B_\mathrm{SZ}$ and $B_\mathrm{X}$ parameters which encode the 
power-law slopes of the scaling relations. We find that the preference for high $B_\mathrm{SZ}$ persists when the X-ray data 
are not used.  In addition, this preference is not localized to any particular region of the data; when considering half the 
cluster sample at a time, either by redshift, $\xi$, or \fsf, the high $B_\mathrm{SZ}$ persists, albeit at a lower significance.

This high $B_\mathrm{SZ}$ implies that the measured observable-space mass function, $dN/d\xi$, is shallower than expected given the 
scaling relation found from the simulations described in \S\ref{sec:sims}. 
We can approximately quantify this by assuming $dN/d\xi$ follows a power law.   The data 
prefers a power-law index of $\sim 4.0$, compared to the simulation prediction of $\sim 5.0$, which 
is disfavored by the measured $dN/d\xi$ at $\sim 4.5\sigma$.   

The preference for a higher $B_\mathrm{SZ}$, i.e., a steeper \zetam\ relation, is statistically weak, approximately $1\sigma$, due to the 
relatively weak $20\%$ width assumed on the $B_\mathrm{SZ}$ prior.  
By contrast,
the assumed width of the prior on the X-ray scaling relation slope $B_\mathrm{X}$ is $5\%$, leading to a tension 
between the likelihood and the prior of approximately $4\sigma$.  There is some evidence for a similarly 
steep \yxm\ scaling relation from the comparison of the V09 and H15 weak lensing mass estimates in Figure~\ref{fig:yxm_norm},
which prefers $B_{X} \sim 0.42$. Such a
slope would be in $\sim4\sigma$ tension with the expected self-similar cluster slope of $B_\mathrm{X}=0.6$.
 
In this work, we will assume that our observational priors on the \yxm\ slope and theoretical priors on the 
mass function slope are well motivated, but note that the shape of the observable-space mass function will 
need to be studied in more detail for future work. 
  
\subsection{$\Lambda$CDM + \summnu}
 
\begin{figure}
\begin{center}
 \includegraphics[angle=0,width=0.47\textwidth]{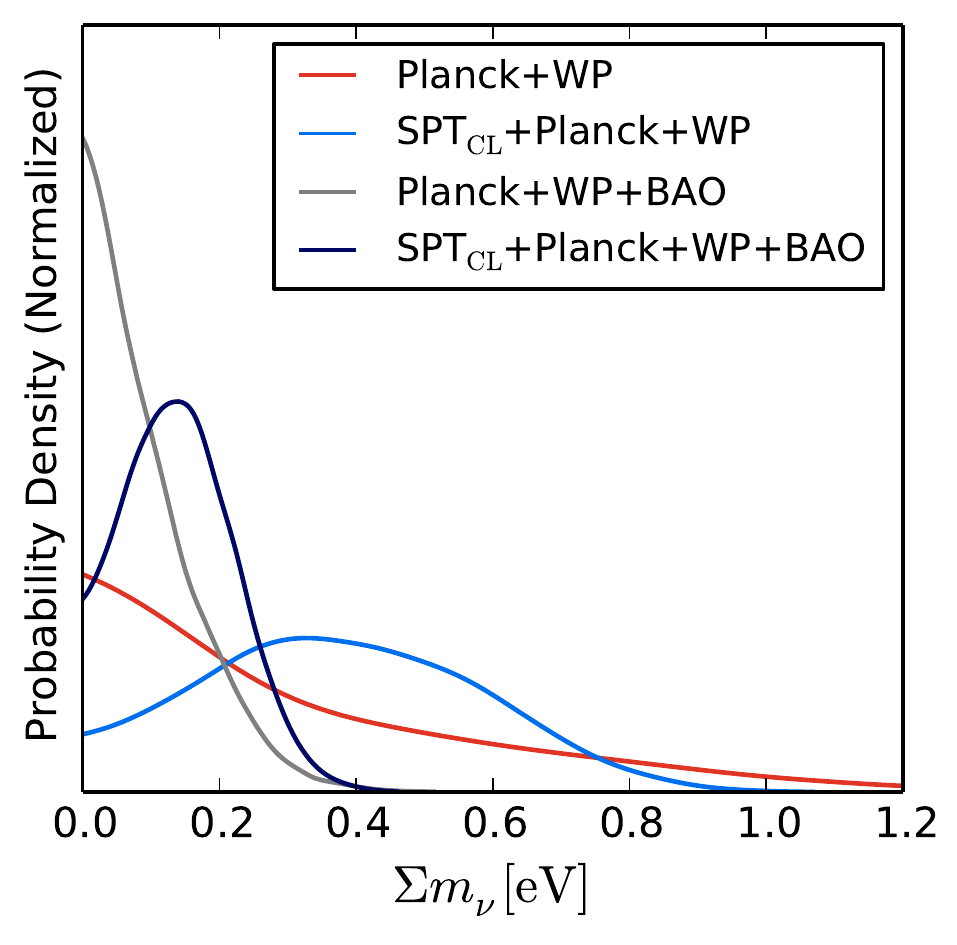}
\end{center}
\caption{Constraints on the species-summed neutrino mass. The addition of cluster constraints to either the \planck+WP or \planck+WP+BAO datasets has a similar effect: the posterior peaks at positive values, but remains consistent with zero.}
\label{fig:mnu_planck_1d}
\end{figure}
 
We now consider a cosmological model in which the species-summed neutrino mass \summnu\ is a free parameter. 
Constraints on this model from the CMB power spectrum show a strong degeneracy between $\sigma_8$ and \summnu\
\citep[e.g.,][]{komatsu09, planck13-16, abazajian15b}.
This allows even modest measurements of $\sigma_8$ to improve on neutrino mass constraints from the CMB power spectrum data alone. 
Figure \ref{fig:mnu_planck_1d} shows the improvement in constraints when adding cluster abundance information 
to CMB power spectrum data alone (\planck+WP), and also when including BAO data.  
In both cases, the addition of cluster information tightens the constraint and 
causes the posterior to peak at positive values of \summnu, though the $95\%$ upper limit 
on \summnu\ remains largely unchanged. 
Allowing the \summnu\ to vary in the range of 0--2~eV, and using the combination of 
the \sptcl+\planck+WP+BAO datasets, we find
\begin{equation}
 \summnu = \mnuplanckbaoresult\ \mathrm{eV}.
 \label{eqn:mnu}
\end{equation}

We note that the preference for positive \summnu, when we combine our cluster abundance measurements
with \planck\ data, is driven by the small residual tension between the preferred values of $\sigma_8$ in
the two datasets. Such a preference has been pointed out by several authors \citep[e.g.,][]{planck13-20,wyman14, battye14}, 
but is in contrast to the preference for positive \summnu\ shown, for example, by the combination of 
\wmap+SPT CMB power spectrum data and SPT cluster data in \citet{hou14}, in which the 
evidence for positive \summnu\ is not driven by the cluster data.
In this work, we find relatively good agreement between the preferred $\sigma_8$ using 
the CMB and \sptcl\ datasets, so the preference for positive \summnu\ is weak and 
consistent at 1$\sigma$ with the minimum expected value of $\sim$0.06 eV from neutrino oscillation experiments 
\citep{lesgourgues06b}.
Relative to the previous cluster-based constraints cited above, the updated weak lensing-based calibration described in \S\ref{sec:wlprior}, 
has shifted the normalization and increased the uncertainty of the observable-mass relation in 
a way that relieves tension with the \planck\ CMB data (see Figure \ref{fig:b13_r13_2500d_14}).  
We also note that our constraint on \summnu\ is largely independent of the change in the optical depth to reionization, $\tau$, between 
the \planck\ 2013 and 2015 data release; the \sptcl\ constraints are independent of $\tau$, and 
the \planck\ CMB constraints on $\sigma_8$ and $\omm$ negligibly changed between the two data releases.

\subsection{$\Lambda$CDM + \summnu\ + \neff}
 
\begin{figure}
\begin{center}
 \includegraphics[angle=0,width=0.47\textwidth]{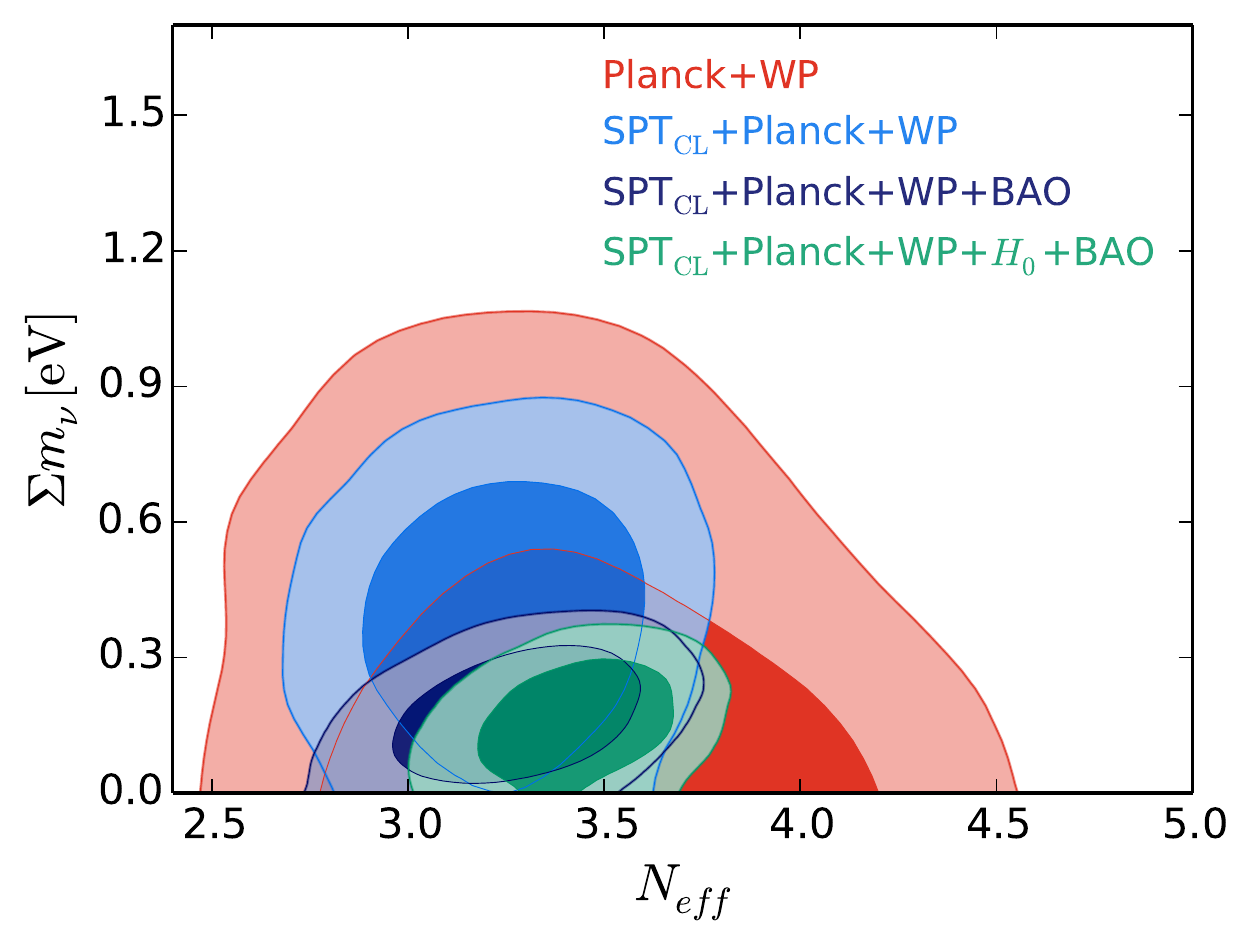}
\end{center}
\caption{Simultaneous constraints on the effective number of relativistic species and the species-summed neutrino mass. The addition of the SPT cluster data reduces the allowed parameter space.}
\label{fig:mnu_nnu}
\end{figure}
 
The effective number of relativistic species, \neff, affects the CMB power spectrum by altering the time of matter-radiation equality, changing the apparent 
sound horizon at recombination \citep[e.g.,][]{hou13}. This mechanism results in strong degeneracies between \neff\ and other cosmological parameters, notably $H_0$ and $\sigma_8$, when considering CMB data alone \citep[e.g.,][]{bashinsky04}. 
Thus, the addition of constraints on $\sigma_8$---such as from the cluster data in this work---and $H_0$ can improve upon CMB-only constraints on \neff.

Here, we 
consider simultaneously varying the species-summed neutrino mass and the effective number of relativistic species. In this cosmological model, the
\planck+WP data alone constrain the $\sigma_8$-\omm-$H_0$ parameter volume relatively poorly.
Adding the cluster information improves on all three of those parameters by roughly a factor of two. Through parameter degeneracies, this improves the simultaneous constraints on \neff\, and \summnu\, as shown in Figure \ref{fig:mnu_nnu}. The simultaneous constraints are
\begin{equation}
 \neff = \nnuresultsptclplanck
\end{equation}
and
\begin{equation}
 \summnu = \mnuresultsptclplancknnufree~\mathrm{eV},
\end{equation}
factors of \nnuresultsptclplanckfracimprovement\ and \mnuresultsptclplancknnufreefracimprovement\ respective improvement over the \planck+WP data alone.
 
Adding BAO data reduces the remaining allowed parameter space significantly to $\neff = \nnuplanckbaoresult$ and $\summnu = \mnuplanckbaoresultnnufree$, and results in a degeneracy between \neff\ and \summnu, allowing for larger values of \neff\ for increasing \summnu.  
These constraints can be further tightened with the addition of local $H_0$ measurements, with the caveat that the best-fit value of $H_0$ from the \sptcl+\planck+WP+BAO dataset,
$H_0 = 68.6 \pm 1.2~\mathrm{km/s/Mpc}$, is in mild tension with direct local measurements from \citet{riess11}. Proceeding to add those local measurements, such that we 
consider a \sptcl+\planck+WP+$H_0$+BAO dataset, we find a preference for larger \neff, resulting in the marginalized constraints of
\begin{equation}
 \neff = \nnuallresult
\end{equation}
and
\begin{equation}
 \summnu = \mnuallresultnnufree~\mathrm{eV}.
\end{equation}
The combined dataset has a \nnuallsignificance\ 
preference for $\neff > 3.046$, the standard model prediction. This is partially
driven by the weak tension between local $H_0$ measurements and the \planck+BAO dataset, as has been noted by other 
authors \citep[e.g.,][]{hou14,wyman14, battye14}. However, the sensitivity to the $H_0$ prior is relatively weak. The preference for \neff\ exceeding the standard model prediction would still be $2.0\sigma$ if the central value of the $H_0$ prior was reduced by one standard deviation.

\subsection{$w$CDM}
\label{sec:wcdm}
  
\begin{figure}
\begin{center}
 \includegraphics[angle=0,width=0.47\textwidth]{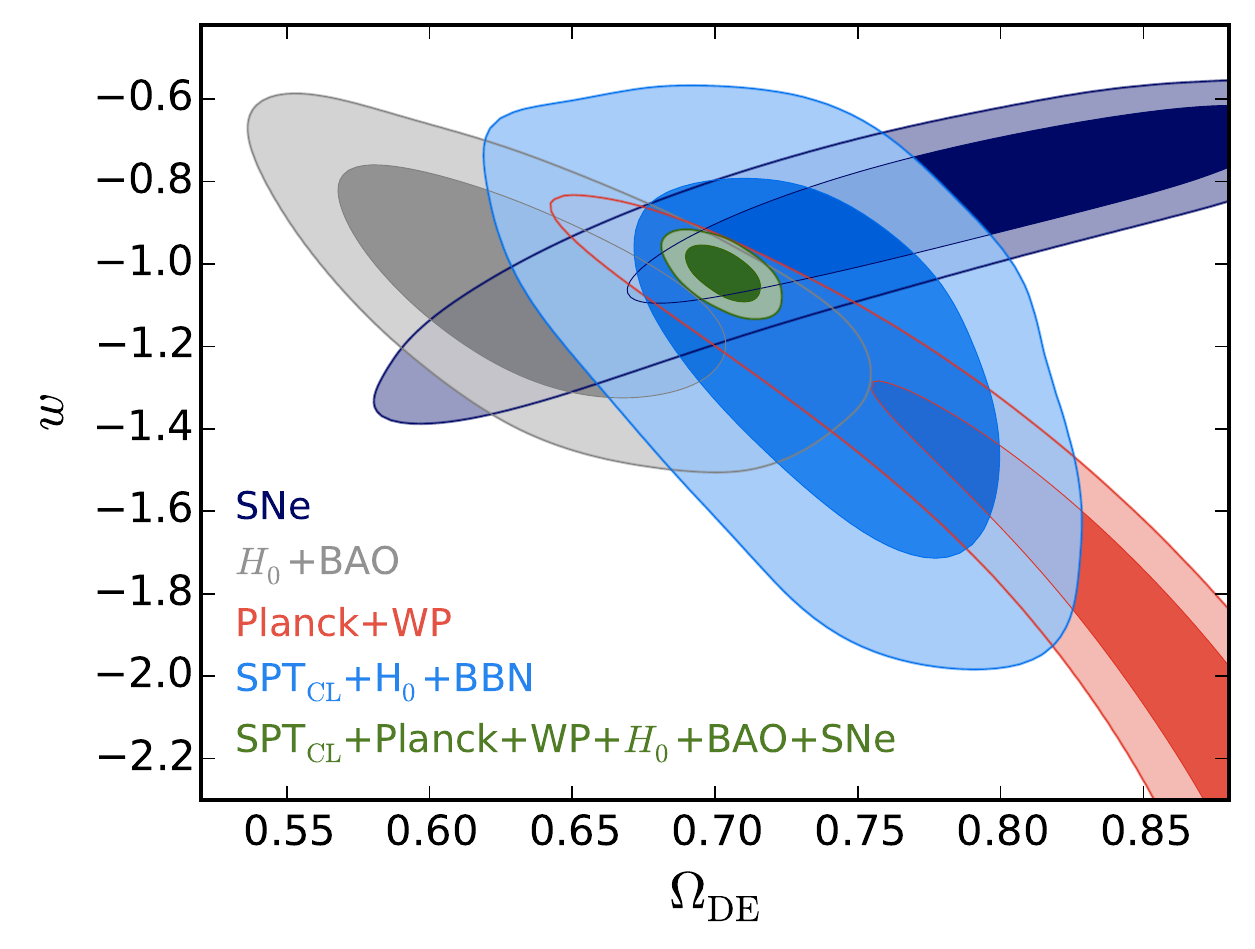}
\end{center}
\caption{Comparison of different cosmological probes of dark energy. The countours show the simultaneous constraints on the present day density of dark energy $\Omega_\mathrm{DE} = 1 - \omm$ and the dark energy equation of state parameter $w$. Using priors on $H_0$ and $\Omega_b h^2$, the SPT cluster data are able to simultaneously constrain the two parameters, and are in good agreement with the other probes. The other probes are sensitive to dark energy primarily through its effect on the geometry of the universe.}
\label{fig:wcdm_2500d_14}
\end{figure}

With the increased number of clusters in this work we are able to place constraints not only on the local cluster abundance but also on the evolution of cluster abundance with redshift.
In particular, we examine the constraints on the \wcdm\ cosmology, where the equation of state of dark energy $w$ is a free parameter. We assume that $w$ is a constant (i.e., its value does not evolve with redshift). This additional parameter affects the cluster abundance and observables through its influence on 
the geometry of the universe and, more importantly, the growth of structure.
The geometrical effects include the change in the survey volume element and the angular diameter distance that modifies the implied X-ray mass information. However, in contrast to other probes of dark energy, the cluster abundance measurement is 
very sensitive to the effect of $w$ on the growth of structure, primarily $\sigma_8(z)$ \citep[e.g.,][]{wang98,haiman01}.

In Figure \ref{fig:wcdm_2500d_14}, we show constraints on the dark energy equation of state parameter $w$ and the energy density of dark energy today $\Omega_\mathrm{DE}$ for different cosmological probes. With the \sptcl+$H_0$+BBN dataset, we obtain
\begin{equation}
 w = \wresultsptclw
\end{equation}
and 
\begin{equation}
 \Omega_\mathrm{DE} = \wresultsptclomegal.
\end{equation}
This is in good agreement, and of comparable precision, with the constraints when considering other 
cosmological probes, including \planck+WP, BAO, and SNe, as can be seen in Figure \ref{fig:wcdm_2500d_14}.  
Since all these probes except the cluster abundance measurement are geometrical tests in this plane, the consistency between the
cluster-implied parameter constraints, and those from other datasets, offers an important systematic test of dark energy. 
This measurement is limited primarily by our knowledge of the redshift evolution of the \zetam\ scaling relation, $C_\mathrm{SZ}$. 
Specifically, both \omm\, and $w$ are correlated with $C_\mathrm{SZ}$ at the $\sim60\%$ level, whereas the correlation 
with $A_\mathrm{X}$ is only $\sim30\%$. 
   
\begin{figure}
\begin{center}
 \includegraphics[angle=0,width=0.47\textwidth]{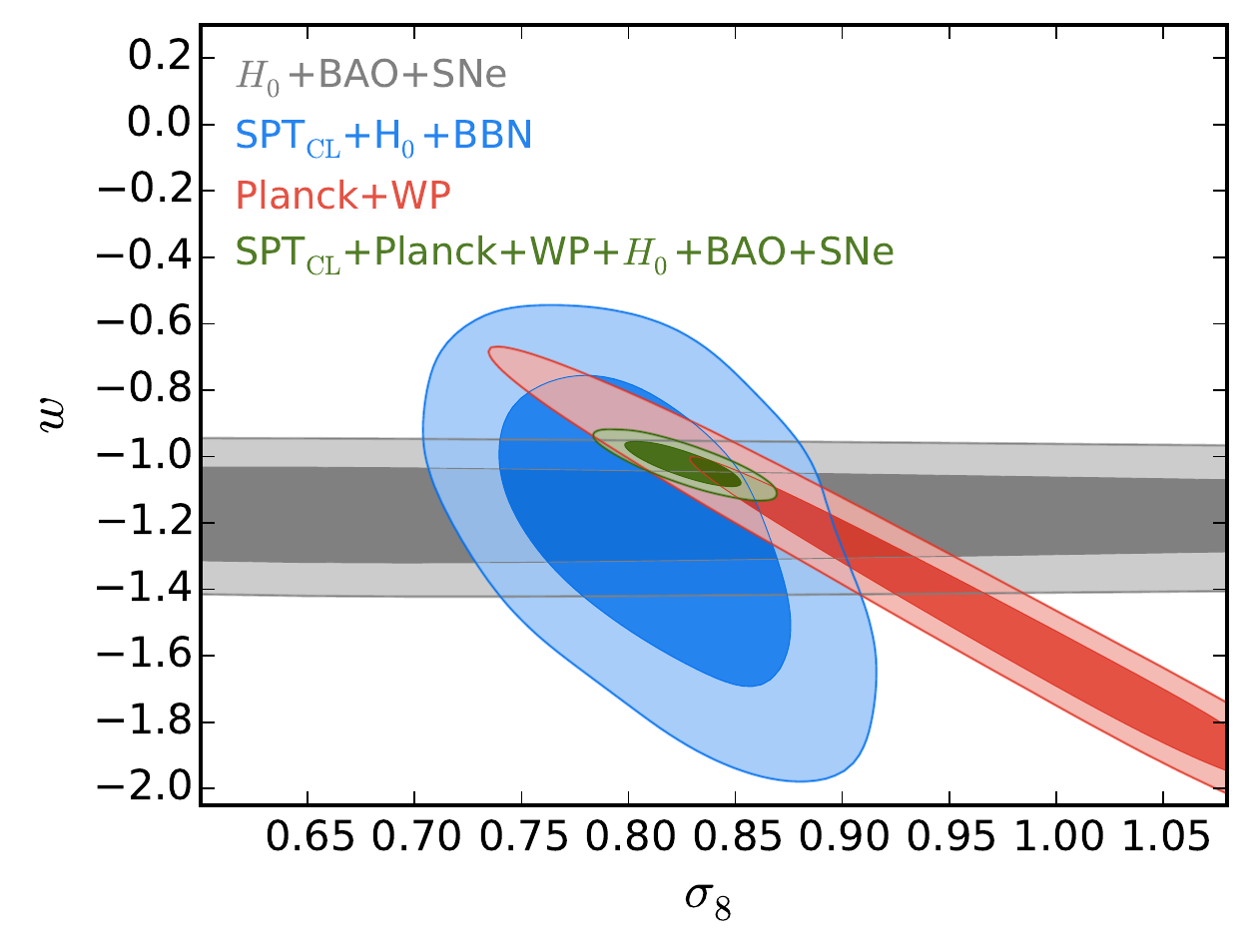}
\end{center}
 \caption{Combined constraints on $w$ and $\sigma_8$. The CMB power spectrum data from \planck+WP shows a strong degeneracy, while the purely geometric constraints from $H_0$+BAO+SNe do not constrain $\sigma_8$. The cluster data simultaneously constrains the two parameters, improving the joint CMB+$H_0$+BAO+SNe constraints both through breaking the $w$-$\sigma_8$ degeneracy present in the CMB constraints, and the direct measurement of $w$.}
\label{fig:w_vs_sigma8}
\end{figure}
 
As shown in Figure \ref{fig:w_vs_sigma8}, the \planck+WP measurements of the primary CMB show a strong degeneracy between $w$ and $\sigma_8$. The addition of cluster data breaks the degeneracy and results in the marginalized constraints
\begin{equation}
 w = \wresultplancksptclw
\end{equation}
and 
\begin{equation}
 \sigma_8 = \wresultplancksptclsigmaeight. 
\end{equation}
This level of $w$-uncertainty is $\sim$2.5 times larger compared to the constraints when adding either of the 
BAO or SNe data sets to the \planck+WP measurements \citep{aubourg14, betoule14}. 

For the dataset combination $H_0$+BAO+SNe, which does not include primary CMB data, $\sigma_8$ is not determined. Adding the cluster data improves the $w$ constraint by \wresulthbaosnefactor\ through a direct measurement of the dark energy parameters $w$ and $\Omega_\mathrm{DE}$, rather than by breaking the $w$-$\sigma_8$ degeneracy. For this combination, we find
\begin{equation}
 w = \wresulthbaosnesptclw.
\end{equation}
Finally, when considering the \planck+WP+$H_0$\allowbreak{+}BAO\allowbreak{+}SNe datasets, $w$ is constrained to \wresultallwnosptcl. The allowed parameter space shows a significant $w$-$\sigma_8$ degeneracy, which allows the addition of the cluster data to improve this constraint by \wresultallfactor, to
\begin{equation}
 w = \wresultallw,
\end{equation}
consistent with \lcdm\ where $w=-1$.
 
\section{Comparison to Other Cluster Surveys}
\label{sec:compare}
 
In this section, we compare the \sptcl\ cosmological constraints to results using other cluster surveys. 
We focus on the \lcdm\ constraints from \S\ref{sec:lcdm}, which employed a \sptcl+$H_0$+BBN dataset, and where 
we constrained $\seight = \constrainthsptclseight$ and $\sigma_8 = \constrainthsptclsigmaeight$.  When comparing to other results, we will discuss 
differences where appropriate.  

Other SZ cluster-based constraints include results from the Atacama Cosmology Telescope \citep[ACT,][]{hasselfield13}
and \planck\ \citep{planck13-20, planck15-24} cluster surveys. However,
comparisons to both are complicated by differences in the assumed mass calibration.  In \citet{hasselfield13}, the ACT collaboration reported  
cosmological constraints using 15 SZ-selected clusters between $0.2 < z < 1.4$.  Several sets of constraints were presented, which assumed 
different priors on the SZ-scaling parameters and also included a calibration based on the dynamical mass measurements from \citet{sifon13}.  The latter relied on a 
scaling relation between velocity dispersion and cluster mass, which was later found to be biased high by $\sim$20\% when using more recent simulations \citep{kirk15}.  
Using a fixed SZ-scaling relation based on the simulations from \citet{battaglia12a} i.e., without including any uncertainty in the 
cluster mass calibration, the \actcl+$H_0$+BBN dataset was used to constrain $\seight = 0.848 \pm 0.032$ and $\sigma_8 = 0.872 \pm 0.065$, consistent with the constraints presented in this work.

The \planck\ collaboration has produced two cluster-based cosmological analyses \citep{planck13-20, planck15-24}. 
\citet{planck13-20} used a sample of 189 SZ-selected clusters between $0.0 < z < 0.55$ with a median redshift 
of 0.15, which is lower redshift than the SPT cluster sample due to the \planck\ selection function.  Assuming an 
identical $H_0$+BBN dataset to our work and assuming a fixed scaling relation except for an overall mass-bias 
factor $b$ with a uniform prior between 0.7 and 1.0, they constrained $\seight = 0.774 \pm 0.024$ and $\sigma_8 = 0.77 \pm 0.03$, 
consistent with our results.
In \citet{planck15-24}, the previous scaling relation calibration was compared to more recent weak lensing 
measurements (H15, WtG), which are also used in this work. For example, H15 found a mass-bias factor of $0.76 \pm 0.08$ for the \planck\ clusters in their lensing sample.
While no numerical constraints were given, 
the newer \planck\ constraints were found to be consistent with the previous \planck\ results, and are also visually in good 
agreement with our results using this mass-bias factor. 

\begin{figure}
 \begin{center}
  \includegraphics[angle=0,width=0.47\textwidth]{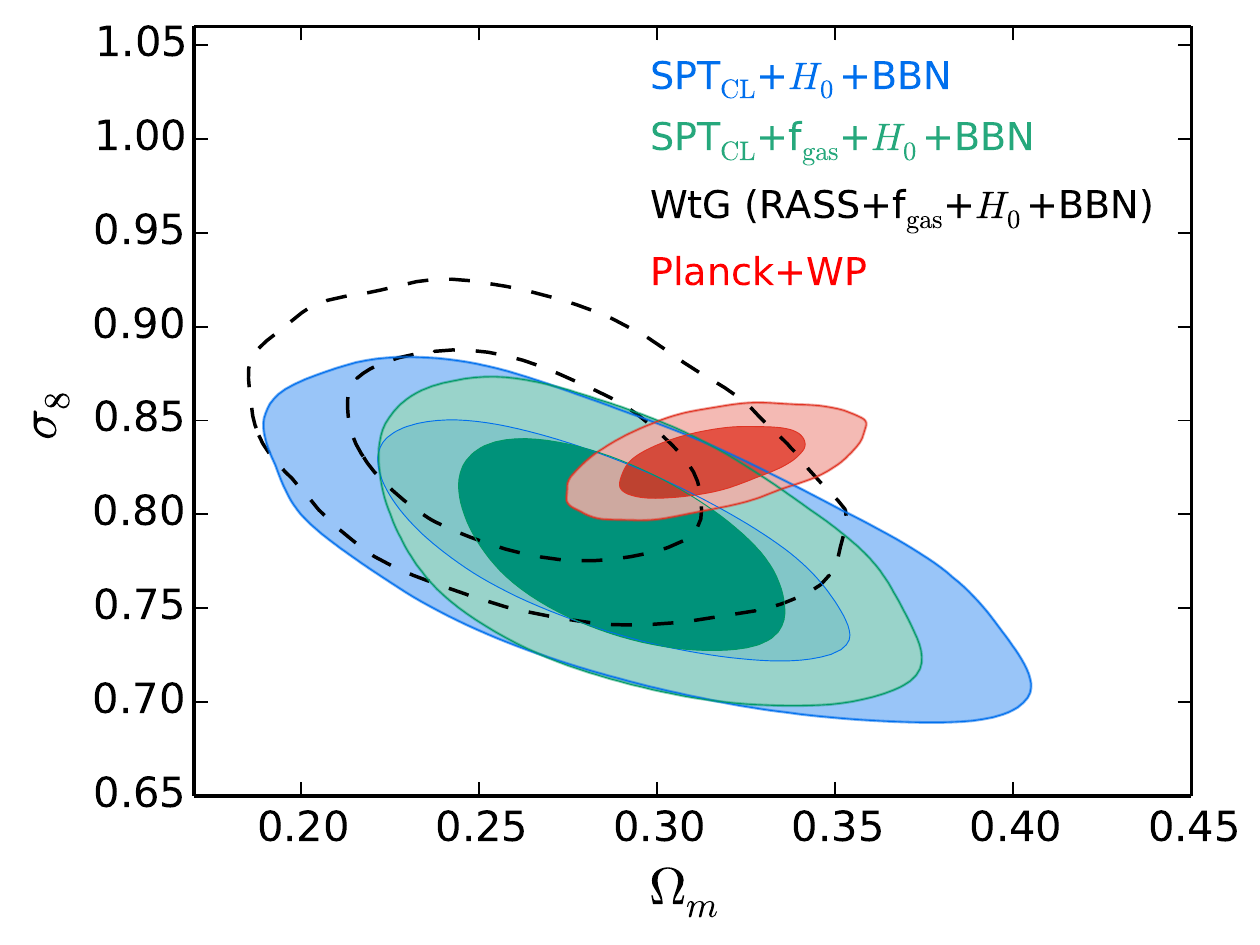}
 \end{center}
 \caption{Comparison of the constraints on $\sigma_8$ and \omm\, from this work, Weighing the Giants (WtG), and \planck+WP CMB measurements. We also show the results from this work after adding the approximate parameter constraints from the $f_\mathrm{gas}$ analysis of \citet{mantz14} for a more direct comparison to the WtG cluster constraints.}
 \label{fig:wtg_comparison}
\end{figure}

In addition, our cosmological constraints are consistent with other previous cluster surveys, including constraints from the 
X-ray selected sample from \citet{vikhlinin09} and the optically selected sample from \citet{rozo10}.

Finally, we compare to more recent cosmological constraints
from WtG \citep{mantz15}.
Their baseline constraints incorporate 
a cluster sample selected from the ROSAT All-Sky Survey (RASS),
follow-up X-ray observations from \chandra, including 
cosmological constraints from cluster gas-fraction \citep[$f_\mathrm{gas}$, see][]{mantz14} measurements, and weak lensing data from WtG; a 
sub-set of which we used to estimate our cluster mass calibration in Section \ref{sec:wlprior}. In Figure \ref{fig:wtg_comparison}, 
we compare directly to the WtG constraints in the $\sigma_8$-$\omm$ plane (WtG find 
marginalized constraints of $\sigma_8 = 0.830 \pm 0.035$ and $\omm = 0.259 \pm 0.030$).  We also plot the \sptcl\ constraints 
with an additional prior of $\Omega_b/\omm h^{1.5} = 0.089 \pm 0.012$, to mimic the
$f_\mathrm{gas}$ constraints.

Overall, we find good agreement between our \sptcl\ dataset and both the \planck-CMB and WtG data.  
The agreement between 
the WtG and \sptcl\ constraints is impressive considering the different selection
methods (SZ vs. X-ray), the different 
X-ray observable (\Yx\ vs. gas mass) and corresponding observable-mass 
scaling relation, as well as 
the independent X-ray analysis pipelines.  
This agreement extends to the $w$CDM cosmological model.  For example, using the WtG cluster data 
(including $f_\mathrm{gas}$ measurements), \citet{mantz15} constrain $w = -0.98 \pm 0.15$, consistent with the 
results in Section \ref{sec:wcdm}.  The combination of the two cluster datasets is 
potentially very powerful for improving cluster-based constraints due to the different redshift ranges of each sample; the 
majority of the WtG sample is at $z < 0.25$, below the lower redshift cut of the \sptcl\ sample used in this work.  This is 
particularly important for cluster-based constraints on dark energy because the combined data sets would provide 
improved constraints on the evolution of the cluster mass function.  This complementarity 
was also noted in \citet{mantz15}, who estimated that the combined datasets could improve the WtG cluster-based 
constraints on dark energy and modified gravity by a factor of $>$2.  

\section{Conclusion}
\label{sec:concl}

In this work, we have taken a well-defined subsample of the SPT cluster catalog from \citet{bleem15b}, selecting only for redshift $z>0.25$ and SPT-SZ detection significance $\xi>5$. 
In order to obtain cosmological constraints, we combine this cluster catalog with 
\chandra\ X-ray observations for \nclustx\ clusters. In addition, we adopt a purely 
weak lensing-based prior on the overall mass scale of the sample from a reanalysis 
of \citet[]{vikhlinin09} using the more recent weak lensing mass 
estimates from \citet[][denoted with H15]{hoekstra15} and the Weighing the Giants 
(WtG) project \citep{kelly14, applegate14, vonderlinden14a, mantz15}. The 1$\sigma$ width of 
this prior is 10\%, which is limited by the small number of clusters in the reanalysis.

The computation of the cluster likelihood uses a new algorithm that scales linearly with the number of mass proxies, where previous algorithms scaled exponentially, which makes incorporating more mass proxies such as weak lensing shear, velocity dispersions, and/or multiple X-ray mass proxies computationally tractable. Our algorithm includes the option to marginalize over all possible correlations between the observables.

Assuming a \lcdm\ cosmology and combining with $H_0$+BBN, we find the marginalized constraints $\sigma_8 = \constrainthsptclsigmaeight$ and $\omm = \constrainthsptclomegam$. The combined parameter combination $\seight$ is constrained to $\constrainthsptclseight$. We find good agreement with the parameter constraints obtained from the WtG project, as well as CMB constraints from either \wmap9+S13 \citep{hinshaw13,story13} or \planck+WP \citep{planck13-16}. We proceed to adopt \planck+WP as our baseline CMB dataset. 

We consider several extensions to the \lcdm\ model. When we allow the species-summed neutrino mass to be a free parameter, the addition of cluster information to CMB information causes the posterior to peak at positive values of neutrino mass (though it is consistent with zero). The same behavior is seen when combining with CMB+BAO, yielding $\summnu = \mnuplanckbaoresult~\mathrm{eV}$.
When further allowing the effective number of relativistic species \neff\ to be a free parameter, and combining with CMB+$H_0$+BAO, we find $\neff = \nnuallresult$ and $\summnu = \mnuallresultnnufree~\mathrm{eV}$. 

Finally, when the dark energy equation of state parameter $w$ is allowed to be free, this cluster catalog can be combined only with priors on $H_0$ and $\Omega_b h^2$ to measure $w = \wresultsptclw$, showing good consistency with the \lcdm\ cosmological model. Adding the cluster data to CMB+$H_0$+BAO+SNe improves the $w$ constraint to $w = \wresultallw$.

The full cosmological power of the 2500~square-degree SPT-SZ cluster survey has not
yet been realized. A joint analysis of the WtG and SPT-SZ cluster samples is currently
being performed. In addition, weak-lensing observations of SPT-SZ 
discovered clusters themselves
are currently being analyzed and are expected to improve on the 10\% mass normalization
uncertainty in this paper, in turn sharpening the cosmological constraints. Especially
important is accurate knowledge of the mass scale over a range of redshifts, which would
specifically improve constraints on models of dark energy or modified gravity. 

The SPT is presently reobserving a 500~square degree patch of the SPT-SZ survey area with the polarization-sensitive receiver SPTpol \citep{austermann12}. 
While the primary science goals are related to polarization, the greater map depth allows for a lower cluster mass threshold, therefore extending the survey out to higher redshift. The next receiver, SPT-3G \citep{benson14}, is currently being built and will allow for significant progress in the SPT cluster program.
The SPT-3G receiver will have a mapping speed $\sim$20~times higher than SPTpol, which should yield $\sim$5000~cluster detections and, importantly, allow cluster mass calibration through CMB-cluster lensing \citep[e.g.,][]{seljak00b, melin14, baxter14}.

\begin{acknowledgements}
 The South Pole Telescope is supported by the National Science Foundation through grant PLR-1248097.  Partial support is also provided by the NSF Physics 
Frontier Center grant PHY-1125897 to the Kavli Institute of Cosmological Physics at the University of Chicago, the Kavli Foundation and the Gordon and 
Betty Moore Foundation grant GBMF 947. This work used resources of McGill University's High Performance Computing centre, a part of Compute Canada. This work was supported in part by the Kavli Institute for Cosmological Physics at the University of Chicago through grant NSF PHY-1125897 and an endowment from the Kavli Foundation and its founder Fred Kavli.  This work was supported in part by the US Department of Energy under contract number DE-AC02-76SF00515. The McGill group acknowledges funding from the National Sciences and Engineering Research Council of Canada, Canada Research Chairs program, and the Canadian Institute for Advanced Research. T.~de~Haan is supported by a Miller Research Fellowship, as well as receiving support from a Natural Science and Engineering Research Council of Canada Postgraduate Scholarship-Doctoral award. 
BB is supported by the Fermi Research Alliance, LLC under Contract No. De-AC02-07CH11359 with the United States Department of Energy.
Argonne National Laboratory’s work was supported under U.S. Department of Energy contract DE-AC02-06CH11357.
D.~Applegate and T.~Schrabback acknowledge support from the German Federal Ministry of Economics and Technology (BMWi) provided through DLR under projects 50 OR 1210 and 50 OR 1407. R.J.F.\ gratefully acknowledges support from the Alfred P.\ Sloan Foundation.
CR acknowledges support from the University of Melbourne and from the Australian Research Council's Discovery Projects scheme (DP150103208).
JHL is supported by NSERC through the discovery grant and Canada Research Chair programs, as well as FRQNT.
The Munich group acknowledges the support of the DFG Cluster of Excellence ``Origin and Structure of the Universe'' and the Transregio program TR33 ``The Dark Universe''. 
The Dark Cosmology Centre is funded by the Danish National Research Foundation.
Optical and infrared followup of SPT Clusters at the Harvard-Smithsonian Center for Astrophysics was supported by NSF grant ANT-1009649.
\end{acknowledgements}
 
\bibliography{}
 
\end{document}